%% file: ms.tex
\documentclass[conference]{ieeeconf}  % Comment this line out if you need a4paper

\IEEEoverridecommandlockouts                              % This command is only needed if 
\title{\LARGE \bf
Prescribed Performance Control Guided Policy Improvement for Satisfying Signal Temporal Logic Tasks
}

\author{Peter Varnai and Dimos V. Dimarogonas$^{1}$% <-this % stops a space
\thanks{This work was partially supported by the Wallenberg AI, Autonomous Systems and Software Program (WASP) funded by the Knut and Alice Wallenberg Foundation, the Swedish Research Council (VR), the SSF COIN project, and the EU H2020 Co4Robots project.}% <-this % stops a space
\thanks{$^{1}$Both authors are with the Division of Decision and Control Systems, School of Electrical Engineering and Computer Science, KTH Royal Institute of Technology, 114 28 Stockholm, Sweden. {\tt\small varnai@kth.se} (P. Varnai), {\tt\small dimos@kth.se} (D. V. Dimarogonas)}%
}

\input{packages}
\input{commandsMain}
\input{commands}
\input{glossary}

\begin{document}

\maketitle
\thispagestyle{empty}
\pagestyle{empty}

%% START OF OWN INCLUDES
%%%%%%%%%%%%%%%%%%%%%%%%%%%%%%%%%%%%%%%%%%%%%%%%%%%%%%%%%%%%%%%%%%%%%%%%%%%%%%%%
\input{abstract.tex}

%%%%%%%%%%%%%%%%%%%%%%%%%%%%%%%%%%%%%%%%%%%%%%%%%%%%%%%%%%%%%%%%%%%%%%%%%%%%%%%%
\input{introduction.tex}

%%%%%%%%%%%%%%%%%%%%%%%%%%%%%%%%%%%%%%%%%%%%%%%%%%%%%%%%%%%%%%%%%%%%%%%%%%%%%%%%
\input{preliminaries.tex}

%%%%%%%%%%%%%%%%%%%%%%%%%%%%%%%%%%%%%%%%%%%%%%%%%%%%%%%%%%%%%%%%%%%%%%%%%%%%%%%%
\input{problem.tex}

%%%%%%%%%%%%%%%%%%%%%%%%%%%%%%%%%%%%%%%%%%%%%%%%%%%%%%%%%%%%%%%%%%%%%%%%%%%%%%%%
\input{solution.tex}

%%%%%%%%%%%%%%%%%%%%%%%%%%%%%%%%%%%%%%%%%%%%%%%%%%%%%%%%%%%%%%%%%%%%%%%%%%%%%%%%
\input{results.tex}

%%%%%%%%%%%%%%%%%%%%%%%%%%%%%%%%%%%%%%%%%%%%%%%%%%%%%%%%%%%%%%%%%%%%%%%%%%%%%%%%
\input{conclusions.tex}

%%%%%%%%%%%%%%%%%%%%%%%%%%%%%%%%%%%%%%%%%%%%%%%%%%%%%%%%%%%%%%%%%%%%%%%%%%%%%%%%
%% END OF OWN INCLUDES

%\addtolength{\textheight}{-12cm}   % This command serves to balance the column lengths
                                  % on the last page of the document manually. It shortens
                                  % the textheight of the last page by a suitable amount.
                                  % This command does not take effect until the next page
                                  % so it should come on the page before the last. Make
                                  % sure that you do not shorten the textheight too much.

%%%%%%%%%%%%%%%%%%%%%%%%%%%%%%%%%%%%%%%%%%%%%%%%%%%%%%%%%%%%%%%%%%%%%%%%%%%%%%%%

%%%%%%%%%%%%%%%%%%%%%%%%%%%%%%%%%%%%%%%%%%%%%%%%%%%%%%%%%%%%%%%%%%%%%%%%%%%%%%%%

%%%%%%%%%%%%%%%%%%%%%%%%%%%%%%%%%%%%%%%%%%%%%%%%%%%%%%%%%%%%%%%%%%%%%%%%%%%%%%%%
%\section*{APPENDIX}
%
%Appendixes should appear before the acknowledgment.
%
%\section*{ACKNOWLEDGMENT}
%
%Any acknowledgments.

%%%%%%%%%%%%%%%%%%%%%%%%%%%%%%%%%%%%%%%%%%%%%%%%%%%%%%%%%%%%%%%%%%%%%%%%%%%%%%%%

\bibliographystyle{IEEEtran}
\bibliography{IEEEabrv,MyBib}

\end{document}

%% file: packages.tex
\usepackage{graphicx, subcaption}

\usepackage{amsmath, amssymb, amsthm}
\usepackage{acro}
\usepackage{cite}

\usepackage{algorithm,algorithmic}
\usepackage{enumerate}
\usepackage{color}

%% file: commandsMain.tex
\newtheoremstyle{bfplain}{}{}{\itshape}{}{\bfseries}{.}{ }{\thmname{#1}\thmnumber{ #2}\thmnote{ (#3)}}
\newtheoremstyle{bfdefinition}{}{}{}{}{\bfseries}{.}{ }{\thmname{#1}\thmnumber{ #2}\thmnote{ (#3)}}
\newtheoremstyle{itdefinition}{}{}{}{}{\itshape}{.}{ }{\thmname{#1}\thmnumber{ #2}\thmnote{ (#3)}}
\theoremstyle{bfdefinition}

\DeclareMathOperator*{\argmin}{arg\,min}

\newcommand{\satisfies}{\vDash}
\renewcommand{\and}{\wedge}
\NewDocumentCommand{\until}{oo}{
	\IfNoValueTF{#1}{\mathcal{U}}{\mathcal{U}_{[#1,#2]}}
}
\NewDocumentCommand{\eventually}{oo}{
	\IfNoValueTF{#1}{\mathcal{F}}{F_{[#1,#2]}}
}
\NewDocumentCommand{\always}{oo}{
	\IfNoValueTF{#1}{\mathcal{G}}{G_{[#1,#2]}}
}

\newcommand{\tp}{^{\textsc{T}}}

\renewcommand{\iff}{\Leftrightarrow}

\newcommand{\bmat}[1]{\begin{bmatrix} #1 \end{bmatrix}}
\newcommand{\subnorm}[1]{_{#1}}
\NewDocumentCommand{\norm}{som}{
	\IfBooleanTF {#1}{\IfNoValueTF{#2} {\| #3 \|}{\| #3 \|\subnorm{#2}}}
	{\IfNoValueTF{#2} {	\left\| #3 \right\|}{\left\| #3 \right\|\subnorm{#2}}}
}
\NewDocumentCommand{\inReal}{soo}{
	\IfBooleanTF {#1}{}{\in}
	\mathbb{R}
	\IfNoValueTF{#2}{}{\IfNoValueTF{#3}{^{#2}}{^{#2 \times #3}}}	
}
\NewDocumentCommand{\inComplex}{soo}{
	\IfBooleanTF {#1}{}{\in}
	\mathbb{C}
	\IfNoValueTF{#2}{}{\IfNoValueTF{#3}{^{#2}}{^{#2 \times #3}}}	
}

\newcommand{\submat}[1]{_{\scriptstyle{#1}}}
\newcommand{\subvec}[1]{_{\scriptstyle{#1}}}
\newcommand{\supmat}[1]{^{\scriptstyle{#1}}}
\newcommand{\supvec}[1]{^{\scriptstyle{#1}}}

\newcommand{\mat}[1]{\mathbf{#1}}
\renewcommand{\vec}[1]{\boldsymbol{#1}}

\newcommand{\defvec}[2] {
	\DeclareDocumentCommand{#1}{s d<> d[] d''} {
		\IfBooleanTF {##1}
			{\IfNoValueTF{##2}{#2}{##2{#2}}}
			{\IfNoValueTF{##2}{\vec{#2}}{##2{\vec{#2}}}}		
		\IfNoValueTF{##3}{}{\subvec{##3}}
		\IfNoValueTF{##4}{}{\supvec{##4}}
	}
}

\newcommand{\defmat}[2] {
	\DeclareDocumentCommand{#1}{s d<> d[] d''} {
		\IfBooleanTF {##1}
			{\IfNoValueTF{##2}{#2}{[##2{#2}]}}
			{\IfNoValueTF{##2}{\mat{#2}}{##2{\mat{#2}}}}			
		\IfNoValueTF{##3}{}{\submat{##3}}
		\IfNoValueTF{##4}{}{\supmat{##4}}
	}
}

\defmat{\A}{A}
\defmat{\B}{B}
\defmat{\C}{C}
\defmat{\D}{D}
\defmat{\E}{E}
\defmat{\F}{F}
\defmat{\G}{G}
\defmat{\H}{H}
\defmat{\I}{I}
\defmat{\J}{J}
\defmat{\K}{K}
\defmat{\L}{L}
\defmat{\M}{M}
\defmat{\P}{P}
\defmat{\Q}{Q}
\defmat{\R}{R}
\defmat{\S}{S}
\defmat{\T}{T}
\defmat{\U}{U}
\defmat{\V}{V}
\defmat{\W}{W}
\defmat{\X}{X}
\defmat{\Y}{Y}
\defmat{\Z}{Z}
\defmat{\NMAT}{0}
\defmat{\SIG}{\Sigma}
\defmat{\LAM}{\Lambda}

\defvec{\ones}{1}
\defvec{\a}{a}
\defvec{\b}{b}
\defvec{\c}{c}
\defvec{\d}{d}
\defvec{\e}{e}
\defvec{\f}{f}
\defvec{\g}{g}
\defvec{\h}{h}
\defvec{\k}{k}
\defvec{\l}{l}
\defvec{\m}{m}
\defvec{\n}{n}
\defvec{\q}{q}
\defvec{\p}{p}
\defvec{\r}{r}
\defvec{\s}{s}
\defvec{\t}{t}
\defvec{\u}{u}
\defvec{\v}{v}
\defvec{\w}{w}
\defvec{\x}{x}
\defvec{\y}{y}
\defvec{\z}{z}
\defvec{\nvec}{0}
\defvec{\lam}{\lambda}
\defvec{\nuvec}{\nu}
\defvec{\pivec}{\pi}
\defvec{\phivec}{\phi}
\defvec{\rhovec}{\rho}
\defvec{\sigvec}{\sigma}
\defvec{\thetavec}{\theta}
\defvec{\alphavec}{\alpha}
\defvec{\gammavec}{\gamma}

%% file: commands.tex
\theoremstyle{bfdefinition}

\newtheorem{problem}{Problem}

\theoremstyle{itdefinition}
\newtheorem{remark}{Remark}

%% file: glossary.tex
\DeclareAcronym{TL}{
	short = TL,
	long = temporal logic
}
\DeclareAcronym{LTL}{
	short = LTL,
	long = linear temporal logic
}
\DeclareAcronym{TLTL}{
	short = TLTL,
	long = truncated linear temporal logic
}
\DeclareAcronym{STL}{
	short = STL,
	long = signal temporal logic
}
\DeclareAcronym{PPC}{
	short = PPC,
	long = prescribed performance control
}
\DeclareAcronym{PI2}{
	short = {PI$^2$},
	long = policy improvement with path integrals
}
\DeclareAcronym{TLPS}{
	short = TLPS,
	long = temporal logic policy search
}
\DeclareAcronym{ReLU}{
	short = ReLU,
	long = rectified linear unit
}
\DeclareAcronym{RL}{
	short = RL,
	long = reinforcement learning
}
\DeclareAcronym{MPC}{
	short = MPC,
	long = model predictive control
}
\DeclareAcronym{MDP}{
	short = MDP,
	long = Markov decision process
}

%% file: abstract.tex
\begin{abstract}

Signal temporal logic (STL) provides a user-friendly interface for defining complex tasks for robotic systems. Recent efforts aim at designing control laws or using reinforcement learning methods to find policies which guarantee satisfaction of these tasks. While the former suffer from the trade-off between task specification and computational complexity, the latter encounter difficulties in exploration as the tasks become more complex and challenging to satisfy. This paper proposes to combine the benefits of the two approaches and use an efficient prescribed performance control (PPC) base law to guide exploration within the reinforcement learning algorithm. The potential of the method is demonstrated in a simulated environment through two sample navigational tasks.

\end{abstract}

%% file: introduction.tex
\section{INTRODUCTION}

\Acp{TL} have gained considerable attention for their convenience and expressive power in specifying complex tasks for a variety of systems. While the field has its roots in formal verification theory \cite{pnueli1977temporal}, recent successful applications include areas in control such as hybrid systems \cite{raman2015reactive}, generating collective swarm behaviors \cite{moarref2017decentralized}, and task and motion planning for robotic systems \cite{saha2018task}. In this paper, we focus on the controller synthesis problem for nonlinear systems subject to tasks specified by \acf{STL}, a type of \acl{TL} originally introduced in the context of monitoring \cite{maler2004monitoring}. In \ac{STL}, the fundamental logical predicates of the language stem from real-valued functions of the system states and the temporal specifications include explicit timing requirements.

\ac{STL} task specifications have lately been studied from a control perspective in the sense of how to ensure their satisfaction. Proposed approaches for controller synthesis include \ac{MPC} \cite{sadraddini2015robust}, \cite{lindemann2016robust}, barrier function- \cite{lindemann2019control}, and \acf{PPC}-based methods \cite{lindemann2017prescribed}. These methods rely heavily on knowledge of system dynamics and exhibit a trade-off between their computational complexity and the range of system dynamics and \ac{STL} task fragments they can handle.

The recent use of \ac{RL} methods in the field of robotics \cite{kober2013reinforcement} and linear temporal logics \cite{sadigh2014learning} have motivated research into their applicability for satisfying \ac{STL} tasks as well. \Ac{RL} is able to deal with unknown dynamics and allows real-time computational expenses to be transferred offline by training from gathered experiences. %Control policies are sought by maximizing a so-called reward function through trial and error, and it is thus crucial for this reward function to properly encapsulate the desired system behavior.
For the purpose of task satisfaction, an \ac{STL} description of the task becomes suitable because \ac{STL} is equipped with various robustness measures that quantify the degree of its satisfaction for a given system trajectory in its entirety \cite{donze2010robust}. Therefore, these measures inherently constitute a descriptive reward to be maximized for task satisfaction and have been shown to be effective for trajectory-based \ac{RL} methods such as \ac{TLPS} \cite{li2018policy}. \ac{TLPS} is based on the \ac{PI2} algorithm \cite{theodorou2010generalized}, which is applicable to continuous state and action spaces and is a form of sampling-based methods also studied for solving \acl{LTL} tasks \cite{fu2017sampling}. The \ac{STL} robustness measures have also been adapted to serve as step-based intermediate rewards for Q-learning in discrete state and action space environments \cite{aksaray2016q}. Practical implementations of \ac{RL} are hindered by the high cost of trial and error (e.g., safety considerations, time-consuming sampling) on which these algorithms generally rely.

%For the purpose of task satisfaction, an \ac{STL} description of the task becomes suitable because \ac{STL} is equipped with various robustness measures that quantify the degree of its satisfaction for a given system trajectory in its entirety \cite{donze2010robust}. Therefore, these inherently constitute a descriptive reward and have been shown to be effective for trajectory-based learning methods such as \ac{TLPS} \cite{li2018policy}. \ac{TLPS} is based on the \ac{PI2} \cite{theodorou2010generalized} algorithm, which is applicable to continuous state and action spaces. The robustness measures have also been adapted to serve as step-based intermediate rewards for Q-learning in discrete state and action space environments \cite{aksaray2016q}. In both cases, practical implementations of \ac{RL} are hindered by the high cost of trial and error (in terms of safety and the time-consuming sampling of real robotic systems) on which the algorithms rely.

The main contribution of this paper is to combine the benefits of model-based \ac{STL} control laws and the reinforcement learning approaches.  More specifically, we propose to use a simple and efficient \ac{PPC} law as a basis for the \ac{PI2} algorithm in order to approximately solve optimal control problems for nonlinear systems subject to \ac{STL} task specifications using partial knowledge of the system dynamics. The learning part allows (locally) optimal solutions to be found under environmental uncertainties, while the base law aids in satisfying the \ac{STL} task and thus leads to effective and robust exploration towards the optimum. The advantages of the approach are illustrated by two simulated scenarios. Although our study is conducted with regards to the trajectory-based \ac{PI2} algorithm, the idea of guided exploration should be applicable to other \ac{RL} methods, such as Q-learning, as well.

The contents of this paper are organized as follows. Section \ref{section:preliminaries} reviews \ac{STL} and the \ac{PPC}-based control law for task satisfaction. A formal problem statement is given in Section \ref{section:problem}. Our main contributions, \ac{PPC} guided policy improvement and practicalities necessary for its effective implementation, are detailed in Section \ref{section:solution}. Section \ref{section:results} presents the simulation study, and we  give concluding remarks in Section \ref{section:conclusions}.

%% file: preliminaries.tex
\section{PRELIMINARIES} \label{section:preliminaries}

\input{preliminaries_system.tex}
\input{preliminaries_stl.tex}
\input{preliminaries_baselaw.tex}

%% file: preliminaries_system.tex
\subsection{System description}

We consider nonlinear systems of the following form:
\begin{equation} \label{eq:systemDynamics}
\x<\dot> = f(\x) + g(\x) \u + \w, \qquad \x(0) = \x[0],
\end{equation}
where $\x \inReal[n]$, $\u \inReal[m]$, and $\w \inReal[n]$ are the state, input, and process noise, respectively. The noise $\w$ is assumed to be zero-mean Gaussian white noise with covariance $\SIG[w] \inReal[n][n]$, while the functions $f(\x)$ and $g(\x)$ are locally Lipschitz continuous with $g(\x)g\tp(\x)$ positive definite for all $\x \inReal[n]$. The system starts in an initial state $\x[0] \inReal[n]$.

A \textit{trajectory} $\tau_{[0, T]}$ of the system (\ref{eq:systemDynamics}) is defined by the signals $\x(t)$ and $\u(t)$ throughout its evolution from $\x[0]$ under the input $\u(t)$ during $t \in [0, T]$. Note that due to the presence of noise, identical initial conditions and inputs may lead to different trajectories. For brevity, we omit the time bounds and simply denote the trajectory $\tau_{[0, T]}$ by $\tau$. Signal values at time $t$ are also often abbreviated as, e.g., $\x[t] := \x(t)$.

%% file: preliminaries_stl.tex
\subsection{\Acf{STL}}

\ac{STL} is a predicate logic defined over continuous-time signals \cite{maler2004monitoring}. The \textit{predicates} $\mu$ are evaluated as true($\top$) or false($\bot$) according to a corresponding function $h^{\mu}:\inReal*[n] \rightarrow \inReal*$ as follows:
\begin{equation}
\mu := \begin{cases}
	\top \text{ if } h^{\mu}(\x) \geq 0, \\
	\bot \text{ if } h^{\mu}(\x) < 0.
\end{cases}
\end{equation}
Predicates can be recursively combined using Boolean logic and temporal operators to form increasingly complex \textit{formulas} $\phi$ (also referred to as task specifications or expressions):
\begin{equation}
\phi := \top \ |\  \mu \ |\ \neg \phi \ |\ \phi_1 \and \phi_2 \ |\ \phi_1 \until[a][b]\phi_2.
\end{equation}
The time bounds of the \textit{until} operator $\until[a][b]$ are given as $a,b \in [0,\infty)$ with $a \le b$. The commonly used temporal operators \textit{eventually} and \textit{always} follow from $\eventually[a][b]\phi = \top \until[a][b]\phi$ and $\always[a][b]\phi = \neg \eventually[a][b] \neg \phi$. A signal $\x(t)$ is said to satisfy an \ac{STL} expression at time $t$ by the following semantics \cite{lindemann2017prescribed}:
\begin{align*}
&(\x, t) \satisfies \mu &&\iff h^{\mu}(\x(t)) \ge 0 \\
&(\x, t) \satisfies \neg\phi &&\iff \neg((\x, t) \satisfies \phi) \\
&(\x, t) \satisfies \phi_1 \and \phi_2 &&\iff (\x, t) \satisfies \phi_1 \and (\x, t) \satisfies \phi_2 \\
&(\x, t) \satisfies  \phi_1 \until[a][b]\phi_2 &&\iff \exists t_1 \in [t+a, t+b] \ : \ (\x, t_1) \satisfies \phi_2\\
& && \quad \ \  \mathrm{and}\ (\x, t_2) \satisfies \phi_1 \  \forall t_2 \in [t, t_1]. 
\end{align*}
The symbol $\satisfies$ denotes satisfaction of an \ac{STL} formula. 

\ac{STL} is equipped with various robustness measures $\rho$ that quantify the extent to which a task specification is satisfied \cite{donze2010robust}. In this work, we employ the so-called \textit{spatial robustness} metric, evaluated as follows for the types of formulas used herein:
\begin{align*}
\rho^\mu(\x, t) &= h^{\mu}(\x(t)) \\
\rho^{\neg \phi}(\x, t) &= -\rho^{\phi}(\x,t) \\
\rho^{\phi_1 \and \phi_2}(\x, t) &= \min\left(\rho^{\phi_1}(\x, t),\rho^{\phi_2}(\x, t)\right) \\
\rho^{\eventually[a][b]\phi}(\x, t)  &= \max_{t' \in [t+a,t+b]}\rho^{\phi}(\x,t') \\
\rho^{\always[a][b]\phi}(\x, t)  &= \min_{t' \in [t+a,t+b]}\rho^{\phi}(\x,t').
\end{align*}

An important property of this robustness metric is that the positiveness of its value indicates whether the corresponding task specification is satisfied or not.

%% file: preliminaries_baselaw.tex
\subsection{\Acf{PPC} for \ac{STL} tasks} \label{subsection:baselaw}

Recent developments aim at designing continuous-time control laws that guarantee the satisfaction of a given \ac{STL} task for a system. In particular, here we review a gradient-based approach advocated by \cite{lindemann2017prescribed} for dynamics of the form (\ref{eq:systemDynamics}). The method uses ideas from \acl{PPC} \cite{bechlioulis2008robust} to guide the robustness metric of logical predicates in time, thereby ensuring their desired temporal behavior. The resulting control law will serve as a basis for guiding learning, as detailed in the upcoming sections of this work.

Consider the following subset of \ac{STL} formulas, defined recursively from predicates $\mu$ as:
\begin{subequations}
	\begin{align}
	\psi &:= \top\ |\ \mu \ | \ \neg \mu \ | \ \psi_1 \wedge \psi_2 \\
	\phi &:= G_{[a,b]} \psi \ | \ F_{[a,b]} \psi \ | \ F_{[a_1,b_1]} G_{[a_2, b_2]} \psi \label{eq:elementarySTL}
	\end{align}
\end{subequations}
where the robustness metric $\rho^{\psi}$ associated with each \textit{non-temporal formula} $\psi$ is assumed concave or convex. The main idea of \ac{PPC} is to achieve satisfaction of the \textit{temporal formula} $\phi$ by controlling the evolution of $\rho^{\psi}$ in time such that it stays bounded between two prescribed curves (a \textit{funnel}) related to the \textit{always} or \textit{eventually} operators. For example, in case of an \textit{always} task, the lower curve remains at or above $0$ during the $[a, b]$ time interval to ensure $\rho^{\psi}(\x(t))\ge0$ and therefore satisfaction of the task $\phi$ as $\rho^{\phi} = \min_{t \in [a,b]} \rho^{\psi}(\x(t)) \ge 0$. We note that the class of satisfiable tasks (\ref{eq:elementarySTL}) could be extended by studying how such funnels should be constructed.

The two prescribed boundaries for $\rho^{\psi}$ are defined by the curve $\gamma(t) \inReal$ and the parameter $\rho_{\text{max}} \inReal$. These are chosen such that the task $\phi$ will be satisfied if $\gamma(t) < \rho^{\psi}(\x(t)) < \rho_{\text{max}}$ holds for all $t \in [a, b]$, i.e., by properly controlling the robustness $\rho^{\psi}$ in time. Under some assumptions, this satisfaction is achieved by the control law:
\begin{equation} \label{eq:stlFeedback}
\u'\phi'(\x, t) = \epsilon(\x, t) g\tp(\x) \dfrac{\partial \rho^{\psi}(\x)}{\partial \x},
\end{equation}
where $\epsilon$ is the so-called transformed error:
\begin{equation} \label{eq:stlTransformation}
\epsilon(\x,t) = S(\xi(\x,t)) \ \text{where} \ \xi(\x,t) = \dfrac{\rho_{\text{max}} - \rho^{\psi}(\x)}{\rho_{\text{max}} - \gamma(t)}.
\end{equation}
Here, the transformation function $S(\xi)$ maps the interval $(0, 1)$ to $(-\infty, \infty)$ in a monotonically increasing manner. This ensures that  $\rho^{\psi}$ stays within its prescribed funnel since $\xi$ approaches 0 or 1 as the upper or lower boundaries are neared. The assumptions require the noise $\w$ to remain in some bounded set $\mathcal{W} \in \inReal*[n]$ and the functions $f(\x)$ and $g(\x)$ to be locally Lipschitz continuous. Furthermore, it must be possible to set the value of the derivative $\dot{\rho}(\x)$ arbitrarily through the input term $g(\x) \u$, such as in case $\rho(\x)$ is concave and $g(\x) g(\x)\tp$ is positive definite for all $\x \inReal[n]$. Note, however, that knowledge of the term $f(\x)$ in the system dynamics (\ref{eq:systemDynamics}) is \textit{not} required to evaluate $\u'\phi'(\x, t)$!

In order to compute the derivative $\partial \rho^{\psi}(\x) / \partial \x$ in the control law (\ref{eq:stlFeedback}), \cite{lindemann2017prescribed} uses a differentiable under-approximation of the robustness associated with the conjunction of propositions: $\rho^{\psi_1 \and \psi_2}(\x,t) \approx -\ln\left(e^{-\rho^{\psi_1}(\x,t)} + e^{-\rho^{\psi_2}(\x,t)}\right)$. This preserves the property that a positive robustness measure $\rho^{\psi}$ implies satisfaction of the corresponding formula $\psi$.

%% file: problem.tex
\section{PROBLEM FORMULATION} \label{section:problem}

The main topic of this paper is summarized by the problem statement given as follows.
\begin{problem} \label{problemStatement}
	Consider the system (\ref{eq:systemDynamics}) starting from an initial state $\x[0] \inReal[n]$ within the time frame $t \in [0, T]$. Design control inputs subject to the constraints $\u \in \mathcal{U}  \subseteq \inReal*[m]$ which guarantee that the system satisfies a given \ac{STL} task $\varphi$ composed of the conjunction of $M$ temporal formulas of the form (\ref{eq:elementarySTL}): \vspace{-2mm}
	\begin{equation} \label{eq:stlForm}
			\varphi = \phi_{1} \wedge \phi_{2} \wedge \ldots \wedge \phi_M,
	\end{equation}
	with a robustness degree of at least $\rho_{\text{min}} \ge 0$ and with respect to minimizing a given cost function $C:\tau \rightarrow \inReal*$ of the generated system trajectory. Only the input term $g(\x)$ is considered known from the system dynamics (\ref{eq:systemDynamics}).
\end{problem}

The cost function $C(\tau)$ indicates preference for one task satisfying trajectory over another, and we assume a solution exists to the outlined optimization problem.
The control inputs are sought over $t \in [0, T]$ in the form of a time-varying policy $\pi_{\theta}(\u[t] | \x[t], t)$ parameterized by $\theta$, which returns the input $\u[t]$ to the system given the current state $\x[t]$ and time $t$.

A similar problem has been formulated and examined for (a broader range of) completely unknown system dynamics in the context of \ac{TLTL}, a language comparable to \ac{STL}, by \cite{li2018policy}. Therein, the goal was to find a policy that maximizes the expected robustness measure corresponding to a general \ac{TLTL} task specification. The authors proposed \acf{TLPS}, a method based on \ac{PI2}, to find such a controller, which was shown to surpass the performance of alternative state-of-the-art algorithms capable of dealing with such a problem. Similar sampling-based methods have also been used for linear temporal logic constrained approximate optimal control \cite{fu2017sampling}.

This work shows that \ac{TLPS} can be further improved by incorporating available knowledge of the system dynamics into the algorithm. Namely, this will be done by using the \ac{PPC} control law introduced in section \ref{subsection:baselaw} to guide \ac{PI2} for an increased rate of convergence and robustness to process noise. We also extend the \ac{PI2} framework to allow optimizing system trajectories subject to \ac{STL} tasks for general $C(\tau)$ costs; task satisfaction is thus treated as a constraint instead of as the target of optimization, in contrast to \cite{li2018policy}. So far, the approach applies to the range of system dynamics (\ref{eq:systemDynamics}) and \ac{STL} formulas (\ref{eq:stlForm}) to which the discussed \ac{PPC} control law is applicable. We intend to extend this range and examine the method's fundamental limitations in future work.   %Furthermore, we take input constraints explicitly into account. These latter modifications are introduced due to their importance in practical applications.

%% file: solution.tex
\section{SOLUTION} \label{section:solution}

The proposed solution to Problem \ref{problemStatement} is based on \acf{PI2}, a trajectory-based \ac{RL} algorithm \cite{theodorou2010generalized}. \ac{PI2} is advantageous in case the system dynamics are (partially) unknown or if the control problem is difficult to solve, e.g., using traditional feedback controllers. This is the case as we aim at both meeting a robustness requirement for satisfying an \ac{STL} formula and minimizing the trajectory cost $C(\tau)$ under input constraints with knowledge of the system dynamics limited to the input term $g(\x)$.

\subsection{The \ac{PI2} framework} \label{section:solution_framework}
Policy improvement finds a control policy $\pi$ under which the generated system trajectory $\tau$ minimizes a given objective function $J(\tau)$\footnote{$J(\tau)$ is a general objective that differs from the target cost $C(\tau)$ introduced earlier and will be defined in Section  \ref{section:solution_robustness}.}. Here, $\pi$ is modeled as a time-varying control policy over a time horizon of length $T$ as:
\begin{equation} \label{eq:basis_control_law}
\pi_{\theta}(\u[t] | \x[t], t) = \u<\hat>(\x[t], t) + \k[t](\theta), \quad t \in [0, T],
\end{equation}
where $\u<\hat>(\x[t], t) \inReal[m]$ is a so-called \textit{base control law} and $\k[t](\theta) \inReal[m]$ is a feedforward term parameterized by the unknown $\theta$. Although $\theta$ could be any low-dimensional parameterization, here we allow degrees of freedom for every time-step in the form $\theta = \{\theta_0, \dots, \theta_T\}$, with each $\theta_t \inReal[n]$. A simple feedforward is then $\k[t] := \theta_t$ as in \cite{li2018policy} or \cite{chebotar2017path}. Here we search for the time differentials of these terms using $\k[t] = \int_{0}^{t} \theta_{\tau} \text{d}\tau$, arguing that the optimal control actions should generally differ marginally from one time instance to another.

The \ac{PI2} algorithm computes a (locally) optimal parameter $\theta$ that minimizes $J(\tau)$ in an iterative fashion, starting from an initial guess $\theta^{(0)}$. The main steps for the $(k)$-th iteration of its variant employed herein are summarized as follows from a combination of the works \cite{li2018policy}, \cite{theodorou2010generalized}, and \cite{stulp2012path}: 
\begin{enumerate}
	\item Generate $i = 1, \dots, N$ samples of parameters $\tilde{\theta}_{t,i} = \mathcal{N}(\theta_{t}^{(k)}, \C[t]^{(k)})$ for each time step $t$ and obtain the system trajectory $\tau_i$ under each corresponding control policy $\pi_{\tilde{\theta_i}}$. The covariances $\C[t]^{(0)} \inReal[n][n]$ are initialized by tuning and will be adapted by the algorithm. Sampling from such Gaussian distributions allows exploration of the parameter space for (locally) optimal policies. 
	\item Compute the cost $J_i = J(\tau_{i})$ of each trajectory $\tau_{i}$ and a corresponding weight $\w*[i]$ using the normalized exponential function:
	\begin{equation} \label{eq:solution_weights}
	\w*[i] = \frac{e^{-\frac{1}{\eta} J_i}}{\sum_{j=1}^{N}e^{-\frac{1}{\eta} J_j}}.
	\end{equation}
	The \textit{temperature parameter} $\eta > 0$ controls the aggressiveness of selecting greedily from the sampled trajectories towards minimizing the objective $J(\tau)$.
	\item Update the policy parameters and apply covariance matrix adaptation using weighted averaging \cite{stulp2012path}:	
	\begin{subequations} 		
		\begin{align} \label{eq:PI2paramUpdate} 
		\hspace{-3mm}\theta_{t}^{(k+1)} &= \sum_{i=1}^{N} \w*[i] \tilde{\theta}_{t,i}, \\ \vspace{-2mm}
		\hspace{-3mm}\C[t]'(k+1)' &=  \C[t,\text{min}] + \sum_{i=1}^{N} \w*[i] (\tilde{\theta}_{t,i} - \theta_{t}) (\tilde{\theta}_{t,i} - \theta_{t})\tp.
		\end{align}
	\end{subequations}
	 The term $\C[t,\text{min}]$ enforces a minimal amount of exploration in subsequent iterations.
\end{enumerate}
The \ac{PI2} algorithm repeats these steps until a given number of K iterations or convergence of $\theta^{(k)}$.

\begin{remark}
	The work \cite{theodorou2010generalized} lays out the theoretical foundation of \ac{PI2} and proves convergence for specific objectives $J(\tau)$. However, the algorithm and its variants are said to perform well even in case the required assumptions do not hold.
\end{remark}

\begin{remark} 
	A significant difference between our outlined algorithm and \cite{theodorou2010generalized} is that here a single cost and weight is assigned to each trajectory in its entirety, in contrast to evaluating a cost-to-go at each time step. This difference was also present in the context of \ac{TLPS} \cite{li2018policy}. Extending the definition of the robustness metric associated with satisfying a given \ac{STL} formula to parts of the trajectory would allow an implementation of \ac{PI2} in its entirety and it would be interesting to see if the result yields any improvements.
\end{remark}

%\begin{remark}
%	Previous work related to temporal logic policy search performed an explicit optimization to update each generated trajectory $\tau_{i}$ towards better minimizing the objective $J_{i}(\tau)$ \cite{li2017policy}; however, the importance of this procedure was not detailed. Policy search has been originally proposed without such an optimization \cite{theodorou2010generalized}, and our resulting algorithm performs well even without such an overhead.
%\end{remark}

In the following, we discuss our choice of the base control law $\u<\hat>$ and a suitable definition of the objective $J(\tau)$ that leads to optimization of the cost $C(\tau)$ subject to a minimum task satisfaction robustness constraint.

\input{solution_baselaw.tex}

\input{solution_robustness.tex}

The definition of the base control law $\u<\hat>(\x, t)$ and the objective $J(\tau)$ defines the \ac{PI2} algorithm outlined in the beginning of this section. The proposed solution for solving Problem \ref{problemStatement} is thus fully summarized as Algorithm \ref{alg:pi2} on the right hand side. (Our experience suggests that for improved convergence and decreased sensitivity to hyperparameters, a Nesterov-type acceleration scheme seen in steps \ref{alg:Nesterov}-\ref{alg:NesterovUpdate} should be employed \cite{nesterov1983method}. This is done by adding momentum towards the direction the solution is currently changing.)

\begin{algorithm} 
\caption{\ac{PPC} guided \ac{PI2} solution to Problem \ref{problemStatement}} 
\label{alg:pi2}
\begin{algorithmic}[1]
	\REQUIRE Initial parameter estimates $\theta^{(0)}_t$, covariances $\C[t]'(0)'$, sample batch size $N$, iteration number $K$, penalty $\lambda$
	\STATE $\alpha^{(0)} := 1$, $\hat{\theta}_t^{(0)} := \theta_t^{(0)} \ \forall t = 0, \dots, T$
	\FOR{$k = 1 \dots K$}
		\FOR{$i = 1 \dots N$}
			\STATE Sample policy parameters $\tilde{\theta}_{t,i} = \mathcal{N}(\hat{\theta}_{t}^{(k-1)}, \C[t]'(0)')$
			\STATE Obtain $\tau_i$ under the \ac{PPC} guided policy $\pi_{\tilde{\theta}_{i}}$ \label{alg:policy}
		\ENDFOR
		\STATE Compute the normalized costs $\bar{J}^{\lambda}_{i}$ for each trajectory $\tau_{i}$ using (\ref{eq:solution_objective}) and (\ref{eq:solution_normalization})
		\STATE Compute weights $w_i$ using (\ref{eq:solution_weights}) with normalized costs
		\FOR{each time step $t = 0, \dots, T$}
			\STATE $\theta_{t}^{(k)} = \sum_{i=1}^{N} \w*[i] \tilde{\theta}_{t,i}$ 
			\STATE $\C[t]'(k)' =  \C[t,\text{min}] + \sum_{i=1}^{N} \w*[i] (\tilde{\theta}_{t,i} - \theta_{t}) (\tilde{\theta}_{t,i} - \theta_{t})\tp$
			%\STATE Apply covariance matrix adaptation (\ref{eq:PI2paramUpdate})
			\STATE $\alpha^{(k)} = (1 + \sqrt{4 \left(\alpha^{(k-1)}\right)^2 + 1})/2$ \label{alg:Nesterov}
			\STATE $\hat{\theta}_{t}^{(k)} = \theta_{t}^{(k)} + (\alpha^{(k-1)} - 1)(\theta_{t}^{(k)} - \theta_{t}^{(k-1)}) / \alpha^{(k)}$
		\ENDFOR
		\STATE Increase penalty term $\lambda$
		\label{alg:NesterovUpdate}
	\ENDFOR
	\RETURN $\theta = \theta^{(K)}$
\end{algorithmic}
\end{algorithm}

%% file: solution_baselaw.tex
\subsection{Base control law} \label{section:solution_baselaw}

The base control law in (\ref{eq:basis_control_law}) is often taken as the linear state feedback $\u<\hat>(\x[t]) = -\K[t] \x[t]$ such as in \cite{li2018policy} and \cite{chebotar2017path}. This choice is general enough to handle cases where the system dynamics are unknown.  However, considering the case where there is partial information available in the knowledge of the input matrix $g(\x)$, we propose to take advantage of the existing \ac{PPC} controller introduced in Section \ref{subsection:baselaw} in order to guide the search procedure towards satisfying the given \ac{STL} task. Using the \ac{PPC} law as a basis for \ac{PI2} offers two main advantages over the linear state feedback. First, it leads to faster convergence of the algorithm, which is important from the practical perspective of sample-efficiency. Second, it can be expected to diminish the algorithm's sensitivity to noise and algorithm hyperparameters. Both of these characteristics will be evaluated in Section \ref{section:results} and are due to the feedback nature of the \ac{PPC} law, as it directly guides system trajectories towards task satisfaction.

For each so-called \textit{elementary} temporal task $\phi_{i}$, $i = 1, \dots, M$ in the task specification (\ref{eq:stlForm}), a corresponding elementary control is defined from (\ref{eq:stlFeedback}) as:
\begin{equation}
\u[\phi_i](\x, t) = -\epsilon_{i}(\x, t) g\tp(\x) \dfrac{\partial \rho^{\psi_i}(\x)}{\partial \x}.
\end{equation}
Since the STL task $\varphi$ is composed of a conjunction of such elementary tasks, the linear combination of these elementary controls can serve well as a base law towards satisfying $\varphi$:
\begin{equation} \vspace{-1mm}
\u<\hat>(\x, t) := \sum_{i=1}^{M} \beta_i \u[\phi_i](\x, t).
\end{equation}
The coefficients $\beta_{i} \inReal$ are such that $\sum_{i=1}^{M} \beta_{i} = 1$. The simulations presented later simply use $\beta_{i} = 1/M$.

While the elementary control laws $\u[\phi_i]$ would individually guarantee the satisfaction of each corresponding task $\phi_{i}$, this cannot be said about their linear combination $\u<\hat>(\x, t)$ with respect to the task $\varphi$. There exist other controllers that can handle such a broader subset of STL tasks \cite{lindemann2019control}; however, these are much more expensive to evaluate than the simple \ac{PPC} law employed herein. We argue that for an algorithm relying on a multitude of sampled trajectories, this computational efficiency makes the \ac{PPC} law a more attractive choice as the base control, because \ac{PI2} will find a task satisfying policy in either case.

It is important to consider the purpose of the base control law from a practical point of view. On the one hand, $\u<\hat>$ should aid \ac{PI2} in finding policies which satisfy the given STL task specification. On the other hand, it must not do so too aggressively in order to allow efficient exploration towards optimizing the cost of interest, $C(\tau)$, using the feedforward terms $\k[t](\theta)$. This suggests that care must be taken when imposing the prescribed boundary curves $\gamma(t)$ and $\rho_{\text{max}}$ that describe the temporal manner in which we aim to satisfy each elementary task $\phi_i$. In future work, we intend on adapting these imposed funnels as the algorithm progresses.

The choice of the transformation function $S(\xi)$ in (\ref{eq:stlTransformation}) is crucial as well, because it determines how aggressively the base law steers the system away from the prescribed robustness boundaries. From a theoretical point of view, the transformation function should grow unbounded as the edges of the funnel are neared to avoid crossing them due to any bounded noise or the unknown term $f(\x)$ in the dynamics. This is achieved by mapping $\xi \in (0, 1)$ to $(-\infty, \infty)$, e.g., using $S(\xi) = -\ln \left((1 - \xi)/\xi\right)$ as in \cite{lindemann2017prescribed}. From a practical perspective within the context of \ac{PI2}, we are not interested in the theoretical guarantee offered by such a choice, as this is lost when taking the conjunction of the elementary controls anyway. Instead, it is more important to avoid numerical issues caused by possibly extremely high values of $S(\xi)$ and its derivative around the interval $\xi \in [0, 1]$. Some of the elementary tasks may not be satisfied in the prescribed manner during exploration; the base control law should still return finite commands in these cases. Exploration also becomes difficult using the feedforward terms if the base control is changing drastically depending on the exact numerical values of the states encountered during discrete-time measurements. Therefore, we propose to use a joined linear and exponential transformation function parameterized by $\alpha > 0$, $\kappa > 0$, $\beta > 0$, and $0 < \xi_c < 1$ that maps $\xi \in [0, 1]$ to $[0, B]$ as:
\begin{equation} \label{eq:functionS}
S(\xi) = \begin{cases}
\max\left(0, \frac{\beta}{\xi_c} \xi\right), \qquad &\xi \le \xi_c \\
m + \alpha (e^{\kappa \xi} - 1), \qquad &\xi > \xi_c.
\end{cases}
\end{equation}
 This construction automatically satisfies $S(0) = 0$ at the prescribed upper robustness boundary $\xi = 0$. The linear part reaches the given value $\beta$ at $\xi_c$, whereas the parameters $\alpha$ and $\kappa$ for the exponential part are chosen such as to make the derivative at the transition $\xi = \xi_c$ continuous and have $S(1) = B$ at the lower robustness boundary $\xi = 1$.

\begin{figure}[hb]
	\centering
	\includegraphics[width=0.95\linewidth,trim={0 0 0 3mm},clip]{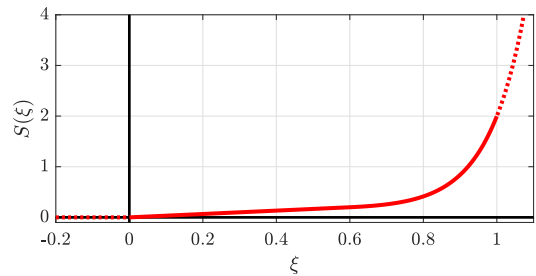}	
	\caption{Example linear-exponential transformation function of the form (\ref{eq:functionS}) defined by $\beta = 0.2$, $B = 2.0$, and $\xi_c = 0.6$.}
	\label{fig:sampleS}
\end{figure}

%% file: solution_robustness.tex
\subsection{Objective function definition} \label{section:solution_robustness}

The objective function $J(\tau)$ plays a central role in \ac{PI2} and must be defined such that the control objectives stated in Problem \ref{problemStatement} are achieved through its minimization. Namely, we wish to find trajectories minimizing the cost $C(\tau)$ while satisfying the \ac{STL} formula $\varphi$ with robustness $\rho_{\text{min}} \ge 0$. 

A common strategy in constrained optimization is to augment the target cost with an appropriate penalty term to enforce the constraint. In our case, we thus have:
\begin{equation} \label{eq:solution_objective}
J^{\lambda}(\tau, \rho) = C(\tau) + P^{\lambda}(\rho),
\end{equation}
where the penalty function $P^{\lambda}(\rho)$ is parameterized by $\lambda \inReal$ and satisfies $P^{\lambda}(\rho \ge \rho_{\text{min}}) \rightarrow 0$ and $P^{\lambda}(\rho < \rho_{\text{min}}) \rightarrow \infty$ as $\lambda \rightarrow \infty$.
Example choices are $P^{\lambda}(\rho) = \rho_0 \cdot e^{\lambda(\rho_{\text{min}} - \rho)}$, or the following function used in this paper:
\begin{equation}
P^{\lambda}(\rho) = \begin{cases}
0 & \rho \geq \rho_{\text{min}}, \\
\lambda (\rho_{\text{min}} - \rho)^3 & \rho < \rho_{\text{min}}.
\end{cases}
\end{equation}
The robustness constraint is enforced by progressively (e.g., linearly or geometrically) increasing $\lambda$ throughout the iterations of \ac{PI2}. Ultimately, setting $\lambda$ to infinity would disregard trajectories that do not satisfy the robustness requirement due to their infinite costs and thus assigned zero weights.

It is important to take into account the practical perspectives related to the objective cost definition. In the original derivation \cite{theodorou2010generalized} of \ac{PI2}, the authors propose a normalization of the $i = 1, \dots, N$ sampled $J^{\lambda}_i$ costs based on their minimal and maximal values:
\begin{equation}
\bar{J}^{\lambda}_i = -h \eta \dfrac{J^{\lambda}_i - \min_j J^{\lambda}_j}{\max_j J^{\lambda}_j - \min_j J^{\lambda}_j},
\end{equation}
where $h$ controls the range of the normalized values, e.g., $h = 10$ used herein. Such cost normalization aims to ensure proper discrimination between the sampled trajectories, which is important for a fast convergence rate of \ac{PI2}.

In our case, some costs $J^{\lambda}_i$ may have extremely high values due to constraint penalization. We thus propose a minor modification to the above formula in order to achieve the desired discrimination. Namely, $\max_j J^{\lambda}_j$ is replaced by a value $J^{\lambda}_{\epsilon}$ for which the $\epsilon \%$ (e.g. $25\%$) of all sampled $J^{\lambda}_i$ costs fall below its value. The normalization thus becomes:
\begin{equation} \label{eq:solution_normalization}
\bar{J}^{\lambda}_i = -h \eta \dfrac{J^{\lambda}_i - \min_j J^{\lambda}_j}{J^{\lambda}_{\epsilon} - \min_j J^{\lambda}_j}.
\end{equation}
This is a more elitist strategy, tuning the normalization for the top $\epsilon$-th percentile of the sampled trajectories and corresponding costs. The normalized cost values are then used to calculate the weights in (\ref{eq:solution_weights}).

%% file: results.tex
\section{RESULTS} \label{section:results}

In this section, we present simulation results of the proposed \ac{PI2} algorithm applied to two sample scenarios. The first involves a simple navigation task with the purpose of illustrating the main advantages of using the \ac{PPC} base law for improved convergence and robustness. The second presents a more complicated scenario to demonstrate the applicability of the technique to a more practical problem. In both scenarios, the feedforward terms of the control policy are parameterized by $\theta = [\theta_0,\ \dots,\ \theta_T]$ with $\k[t] = \sum_{t' = 0}^{t} \theta_{t'}$, the discrete form of the expression described in Section \ref{section:solution_framework}. As a basis for comparison, Algorithm 1 is also implemented using the linear state feedback $\u<\hat>(\x[t]) = -\K[t] \x[t]$ as the base law in step \ref{alg:policy}; the two variants are referred to as the `LIN' and `PPC' variants, respectively.

\input{results_simple.tex}

\input{results_complex.tex}

%% file: results_simple.tex
\subsection{Simple navigation task}

Consider an omnidirectional robot described by the simplified dynamics $\x<\dot> = [\dot{x}\ \dot{y}]\tp = [u_x\ u_y]\tp = \u$ and limited by the input constraint $\norm[2]{\u} \le 1$. The robot is initially located at $\x[0] = [3.0\ 0.3]\tp$. 

The robot is tasked to navigate within an $r_g = 0.2$ radius goal region at $\x[g] = [1.0\ 3.5]\tp$ within 10 seconds while avoiding a large circular obstacle of radius $r_o = 1.2$ centered at $\x[o] = [2.5\ 2.0]\tp$.  We aim for a robustness of at least $\rho_{\text{min}} = 0.05$, as well as to minimize the time this is first attained for the subtask $\psi_1$ of reaching the goal region, i.e., $C(\tau) = \argmin_{t}\{t: \rho^{\psi_1}(t) \ge \min\left(\rho_{\text{min}}, \max_t \rho^{\psi_1}(t)\right)\}$. The minimum between $\rho_{\text{min}}$ and $\max_t \rho^{\psi_1}(t)$ is taken in order to define an appropriate cost for the case when $\psi_1$ is not yet satisfied with the desired robustness. The scenario is simulated for $T = 10s$ with resolution $\Delta t = 0.05s$.

The formal \ac{STL} specification of the task is $\varphi = \phi_1 \and \phi_2 = \eventually[0][10] \psi_1 \and \always[0][\infty]\psi_2$ where $\psi_1 = (r_g - \norm{\x - \x[g]} > 0)$ and $\psi_2 =  (\norm{\x - \x[o]} - r_o > 0)$. The corresponding funnels enforcing these temporal behaviors are described by the parameters $\rho_{\text{max}}$, $\gamma_0$, $\gamma_{\infty}$, and $t_c$: the upper boundary is $\rho_{\text{max}}$, while the lower one increases linearly from $\gamma_0$ to $\gamma_{\infty}$ within $t_c$ seconds and remains at that value thereafter. For the two subtasks, the parameters were $\rho_{\text{max}} = \{r_g, 1.0\}$, $\gamma_0 = \{{-}4.0, \rho_{\text{min}}\}$, $\gamma_{\infty} = \{\rho_{\text{min}}, \rho_{\text{min}}\}$, and $t_c = \{10, 10\}$ (the $i$th element in each set refers to the values used for the subtask $\phi_i$). The parameters used to define the corresponding linear-exponential transformation functions (\ref{eq:functionS}) are given as $B = \{2.0, 2.0\}$, $\beta = \{0.8, 0.1\}$, and $\xi_c = \{0.5, 0.8\}$; the first one pulls consistently towards the goal, while the second one pushes away from the obstacle mainly when it is nearby.

\begin{figure}[b]
	\centering
	\begin{minipage}[c][0.85\linewidth][t]{\linewidth}
		\centering
		\begin{subfigure}{0.5\linewidth}
			\centering
			\includegraphics[width=.8\linewidth,trim={0 0 0 0},clip]{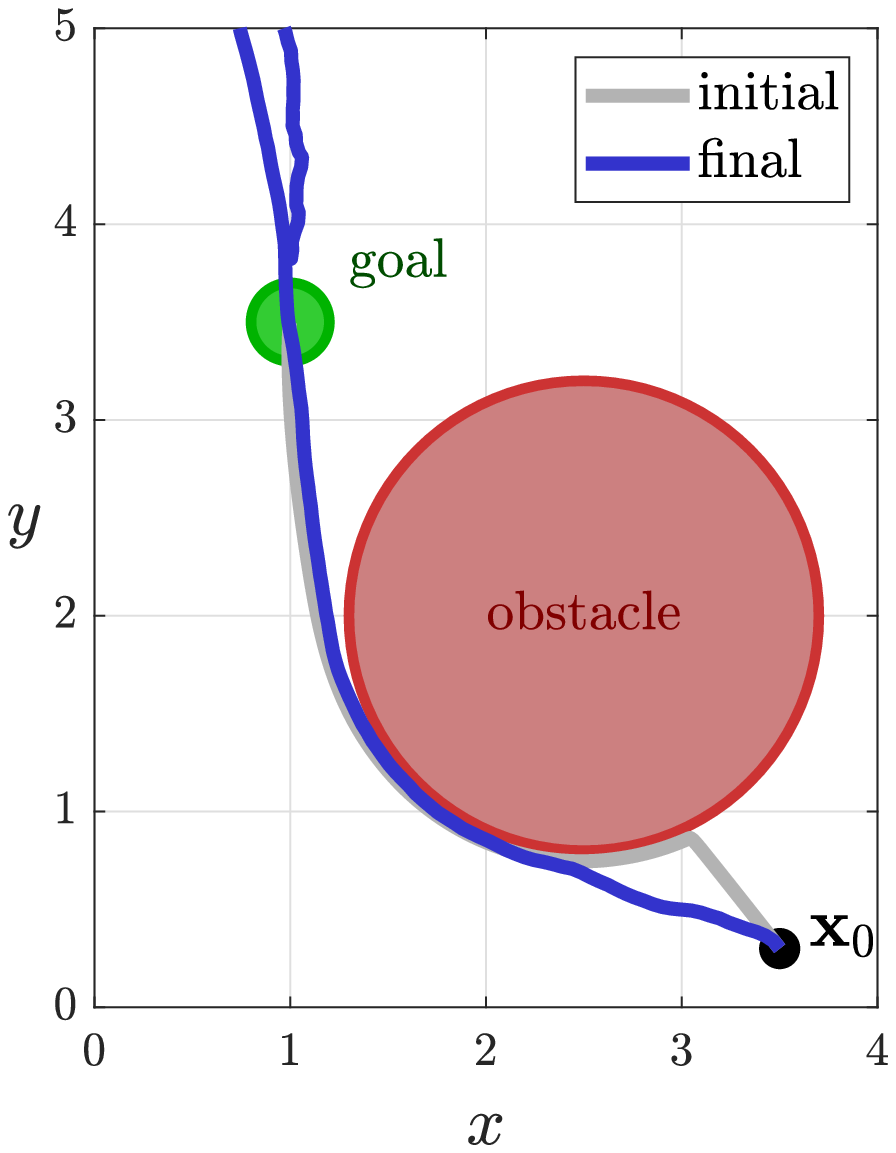}
			\caption{}
			\label{fig:scenario1_sampleResult}
		\end{subfigure}%
		\begin{subfigure}{0.5\linewidth}
			\centering
			\includegraphics[width=.8\linewidth,trim={0 0 0 0},clip]{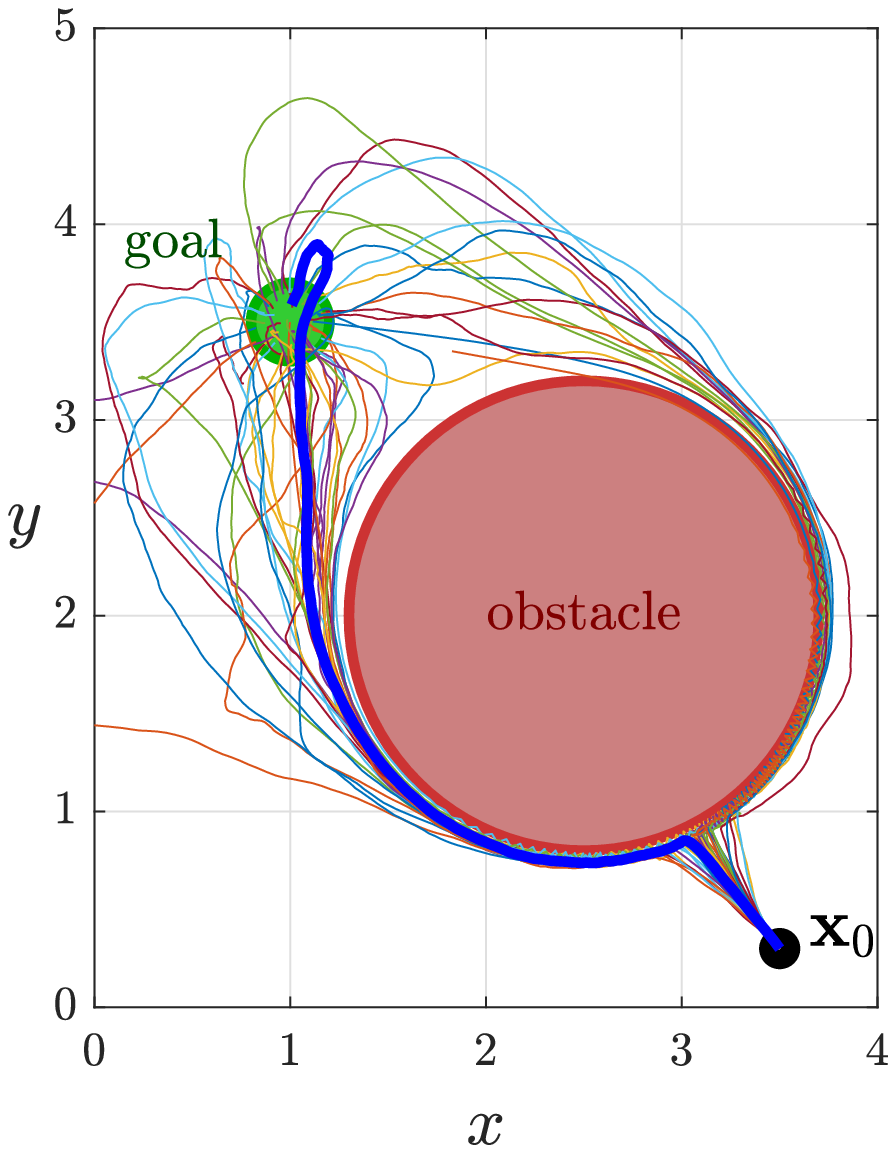}
			\caption{}
			\label{fig:scenario1_sampleIteration}
		\end{subfigure}
		\caption{(a) Initial and obtained trajectories for the simple navigation task scenario using \ac{PPC} guided \ac{PI2}. The result achieves cost $C(\tau) = 4.50$ with robustness $\rho^{\varphi} = 0.046$ for task satisfaction. (b) Visualization of the first iteration of the underlying \ac{PI2} algorithm. 50 trial trajectories are shown, along with the iteration's obtained result highlighted in blue.}
		\label{fig:scenario1}		
	\end{minipage}
\end{figure}

The hyperparameter values used in Algorithm \ref{alg:pi2} were $K=25$, $N = 100$, $\epsilon = 25\%$, $\C[t,0] = 2 \cdot 10^{-3} \I[2]$, and $\C[t,\text{min}] = 2 \cdot 10^{-4}\I[2]$ for all time steps, where $\I[n]$ denotes the identity matrix of size $n$. To enforce the \ac{STL} task constraint, the penalty parameter $\lambda$ was increased logarithmically from 2 to 2000 throughout the $K$ iterations. The navigation task is depicted in Fig. \ref{fig:scenario1_sampleResult}, along with results from \ac{PPC} guided \ac{PI2}.

\begin{figure*}[t]
	\centering
	\begin{subfigure}{0.33\textwidth}
		\centering
%		\includegraphics[width=.9\linewidth]{figures/scenario1_comparisonACC_J.eps}
%		\smallbreak
		\includegraphics[width=.9\linewidth,trim={0 0.85cm 0 0},clip]{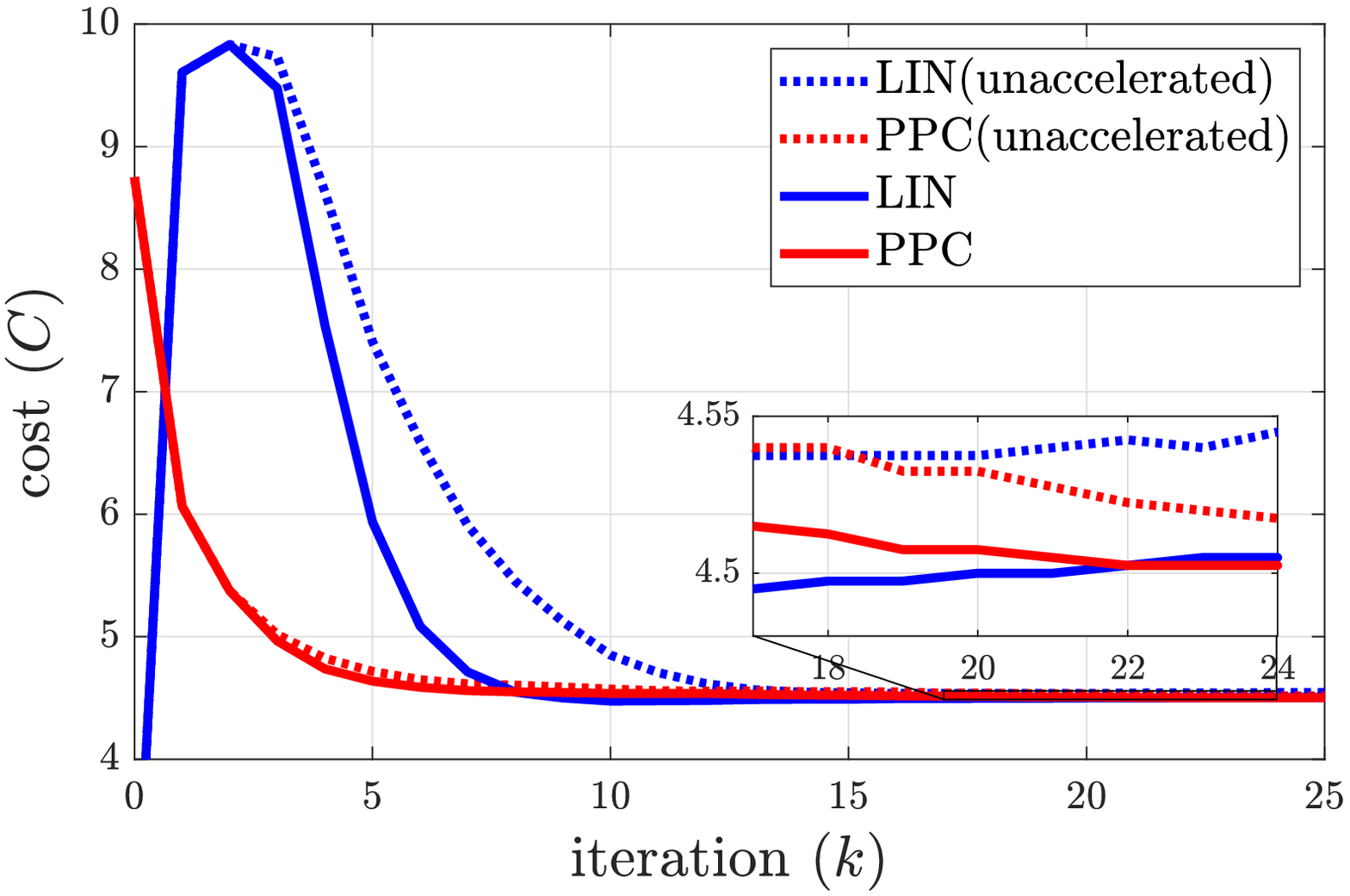}
		\smallbreak
		\includegraphics[width=.9\linewidth]{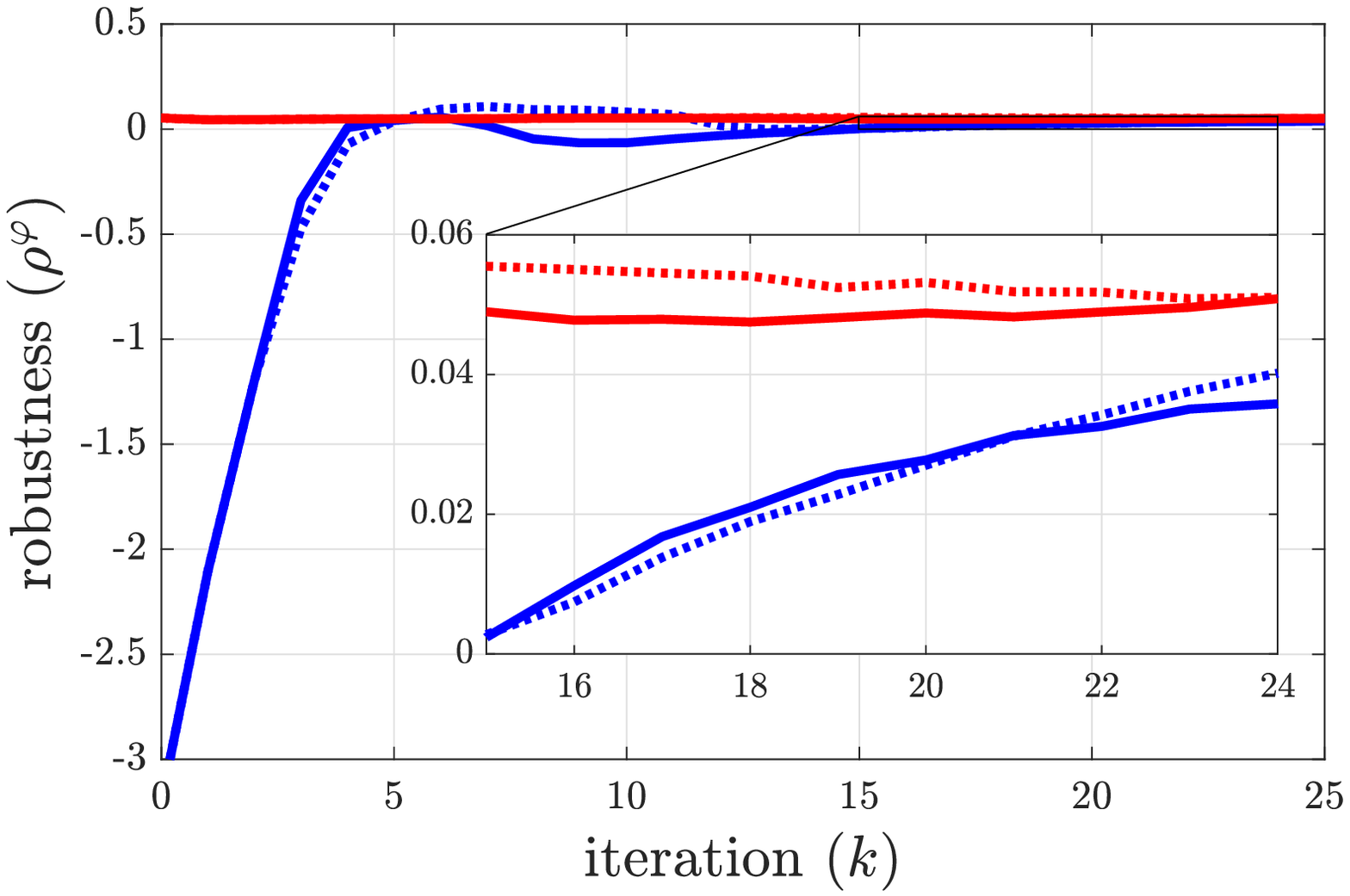}
		\caption{Employing Nesterov-type acceleration}
	\end{subfigure}
	\begin{subfigure}{0.33\textwidth}
		\centering
%		\includegraphics[width=.9\linewidth]{figures/scenario1_comparisonN_J.eps}
%		\smallbreak
		\includegraphics[width=.9\linewidth,trim={0 0.85cm 0 0},clip]{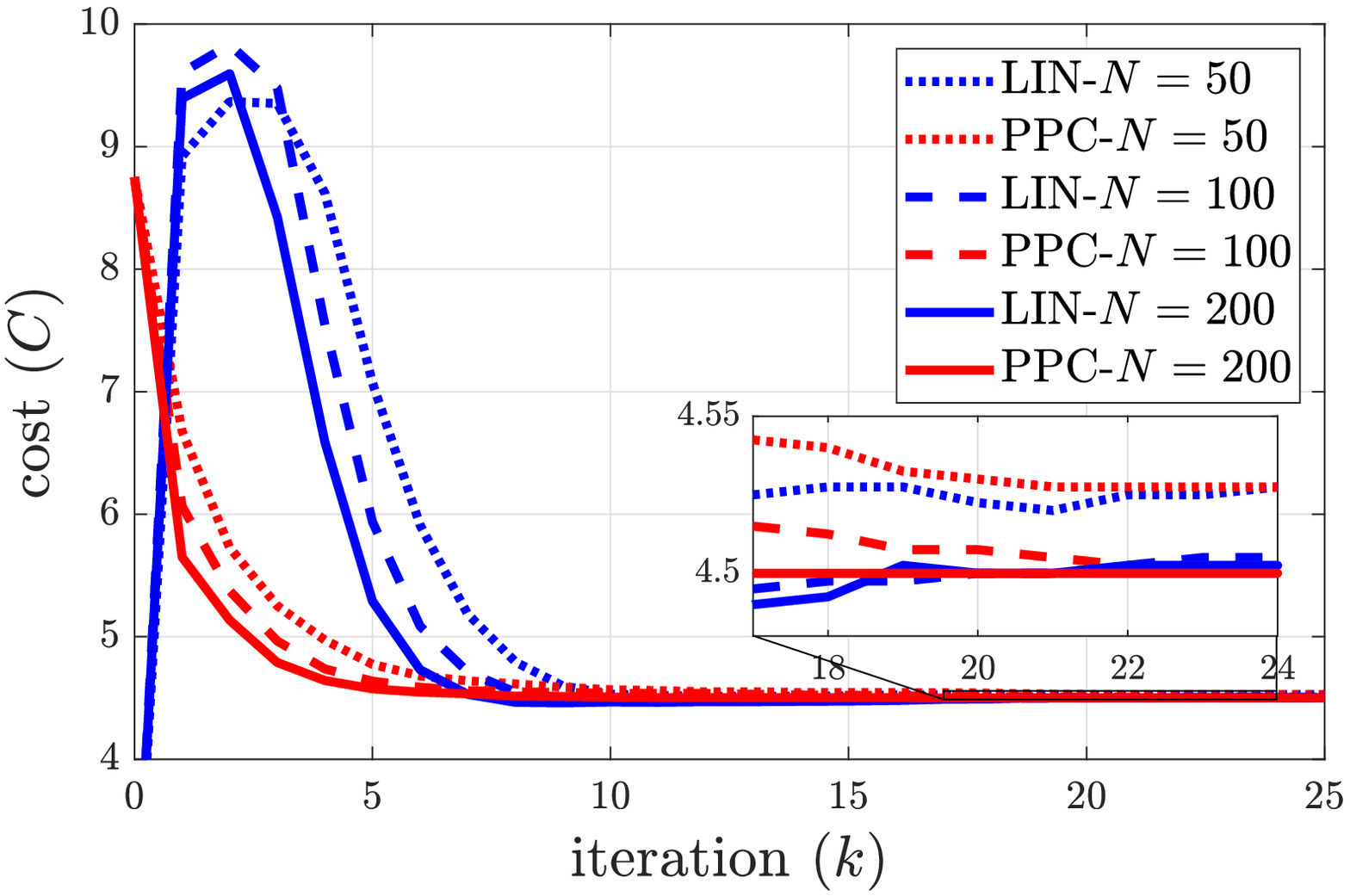}
		\smallbreak
		\includegraphics[width=.9\linewidth]{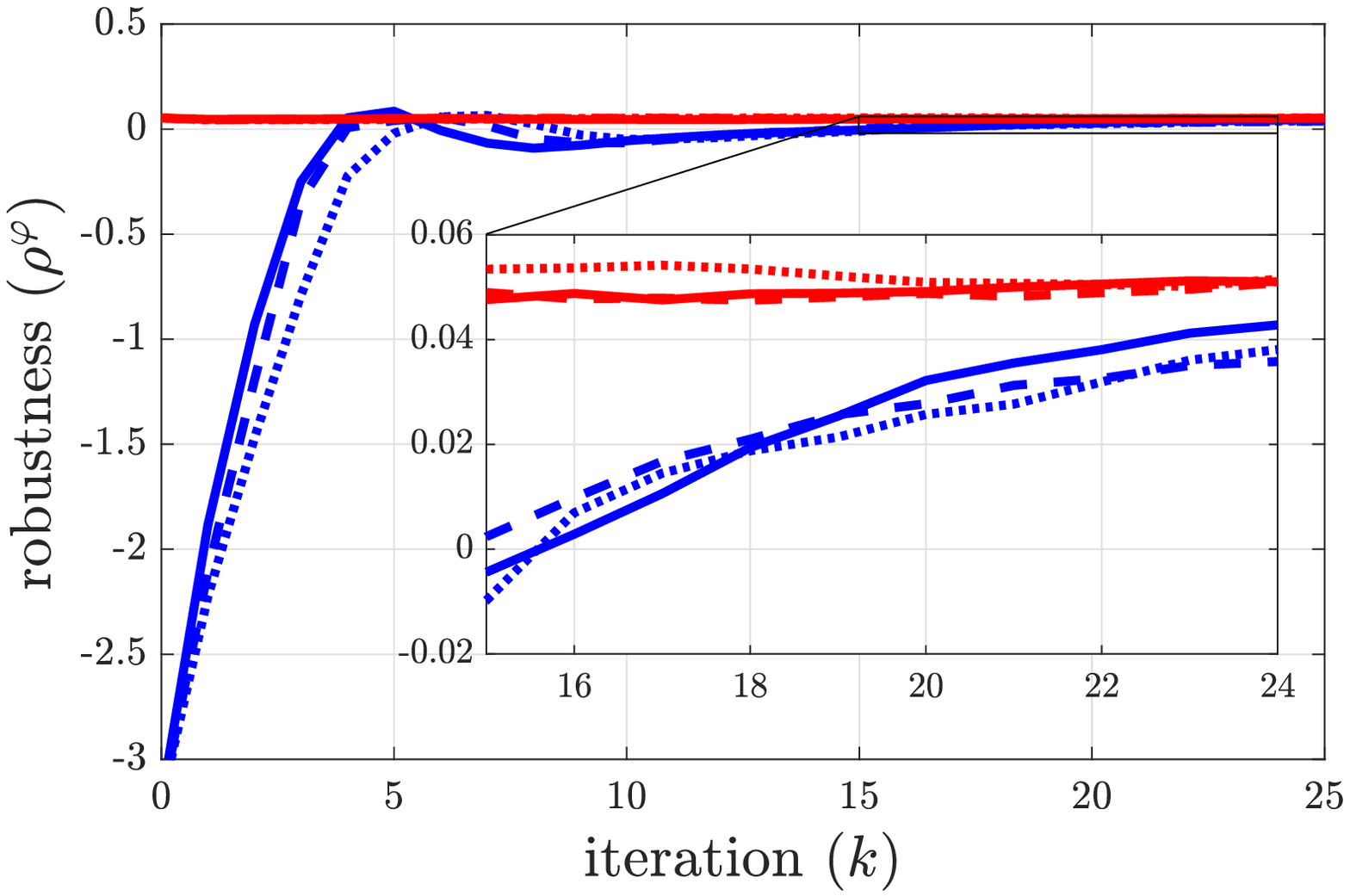}
		\caption{Sample size $N$}
	\end{subfigure}%
	\begin{subfigure}{0.33\textwidth}
		\centering
%		\includegraphics[width=.9\linewidth]{figures/scenario1_comparisonCmin_J.eps}
%		\smallbreak
		\includegraphics[width=.9\linewidth,trim={0 0.85cm 0 0},clip]{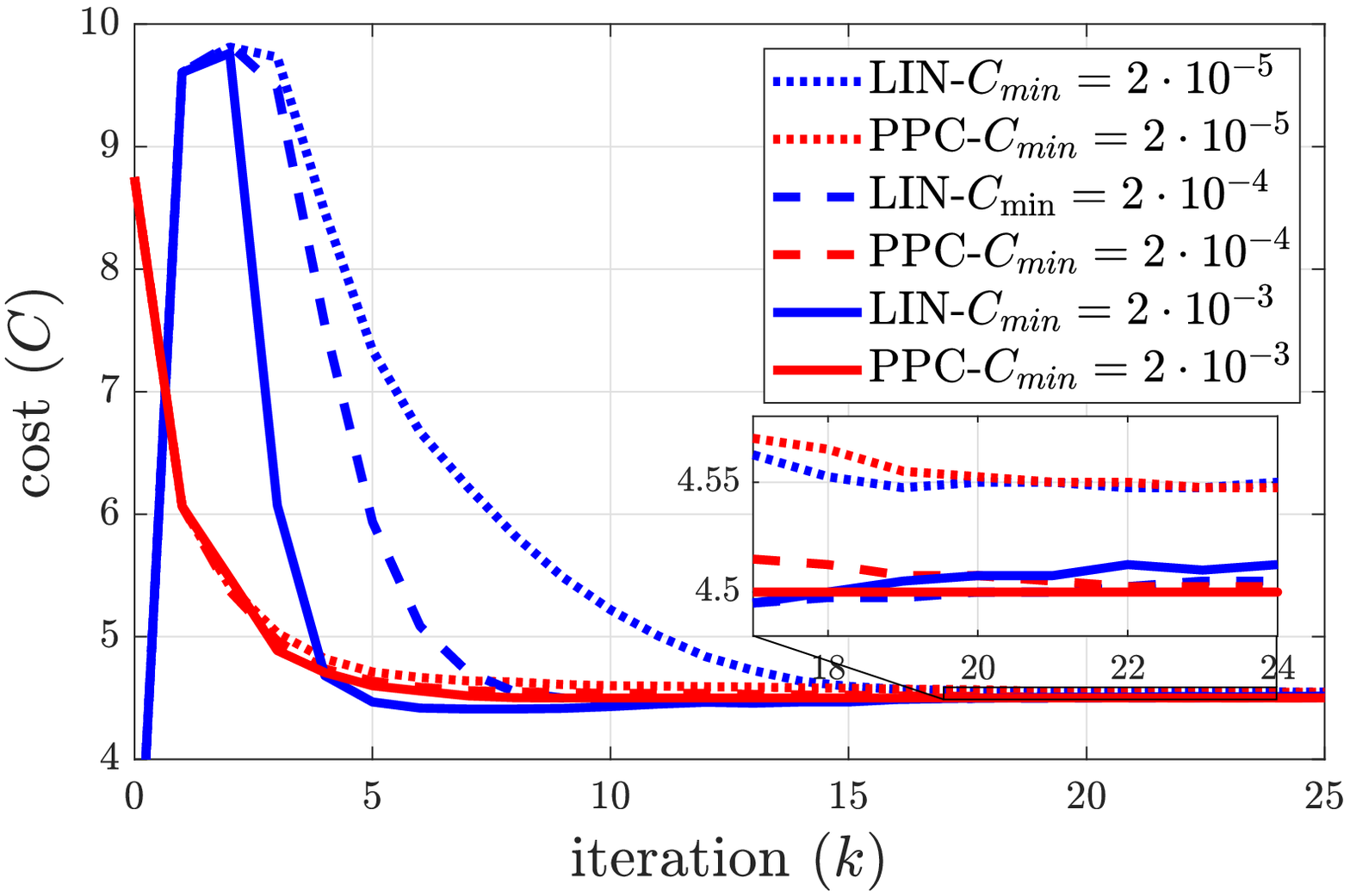}
		\smallbreak
		\includegraphics[width=.9\linewidth]{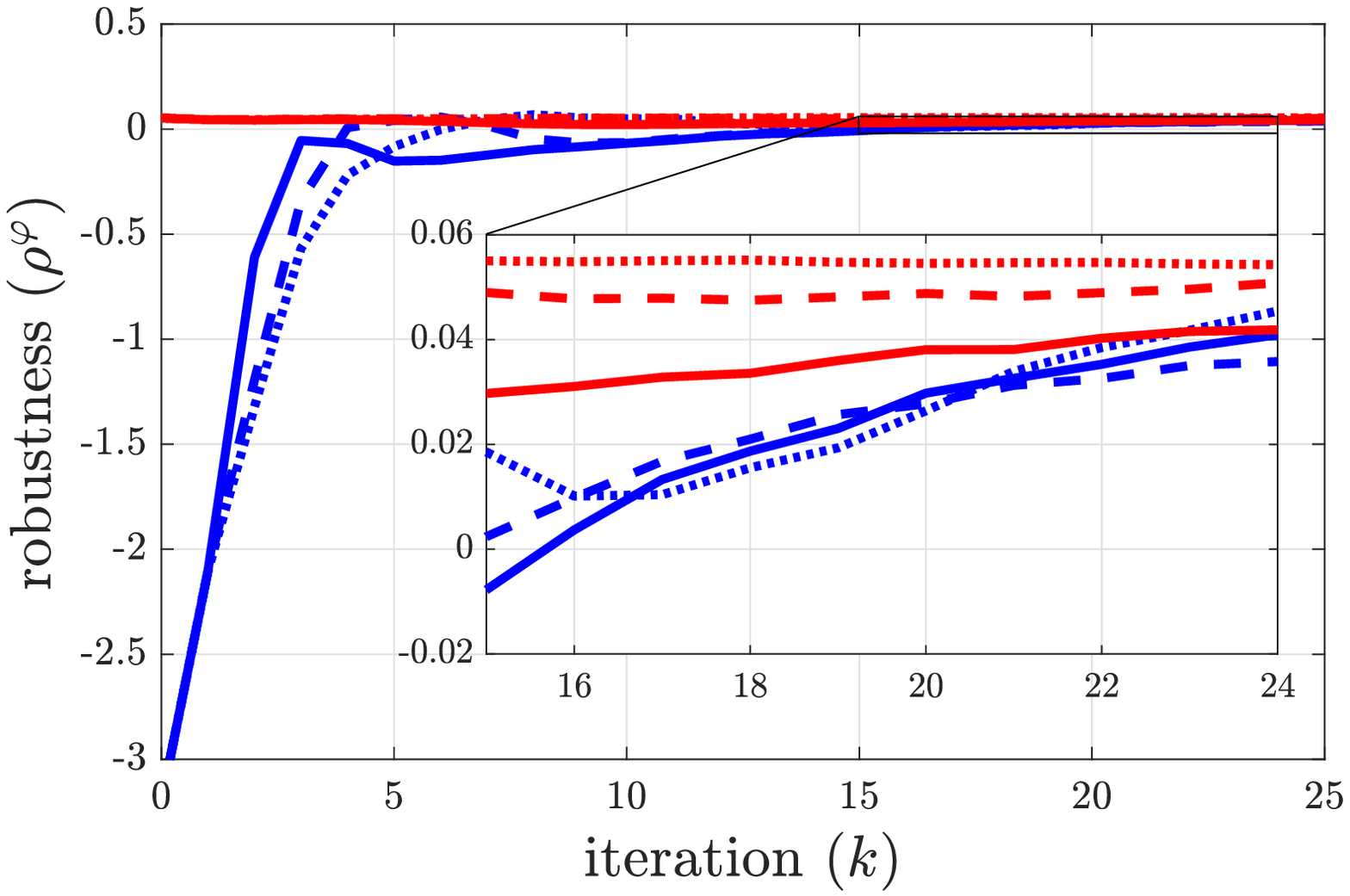}
		\caption{Minimum covariance $C_{\text{min}}$}
	\end{subfigure}%
	\caption{Effect of two chosen \ac{PI2} hyperparameters and Nesterov's acceleration scheme on the algorithm's performance. The results show an average of $20$ sample runs in each case. The \ac{PPC} base law aids task satisfaction and allows for efficient exploration directly towards the cost $C(\tau)$ of interest, achieving faster convergence with less sensitivity to hyperparameters.}
	\label{fig:scenario1_convergenceComparison}
\end{figure*}

We first examine the scenario without the presence of any process noise. Fig. \ref{fig:scenario1_convergenceComparison} on the next page compares the convergence rate of Algorithm \ref{alg:pi2} between using linear state feedback (with $\K[t] = \I[2]$) and the described \ac{PPC} law as a basis control law. The graphs were obtained by varying different hyperparameters of the algorithm and averaging 20 randomized runs for each case. It is clear that the \ac{PPC} law outperforms its linear feedback-based counterpart both in terms of improved convergence rate and lower sensitivity to the examined parameters. Applying the Nesterov acceleration scheme leads to improvements in both cases, though the difference is less evident for the \ac{PPC} variant due to the simplicity of the navigation task.

\begin{figure}[b]
	\centering
	\begin{minipage}[c][0.85\linewidth][t]{\linewidth}\centering
		\captionsetup[subfigure]{aboveskip=1.5mm}
		\begin{subfigure}{0.5\linewidth}		
			\centering
			\includegraphics[width=\linewidth,trim={2cm 0.2cm 2.4cm 0.05cm},clip]{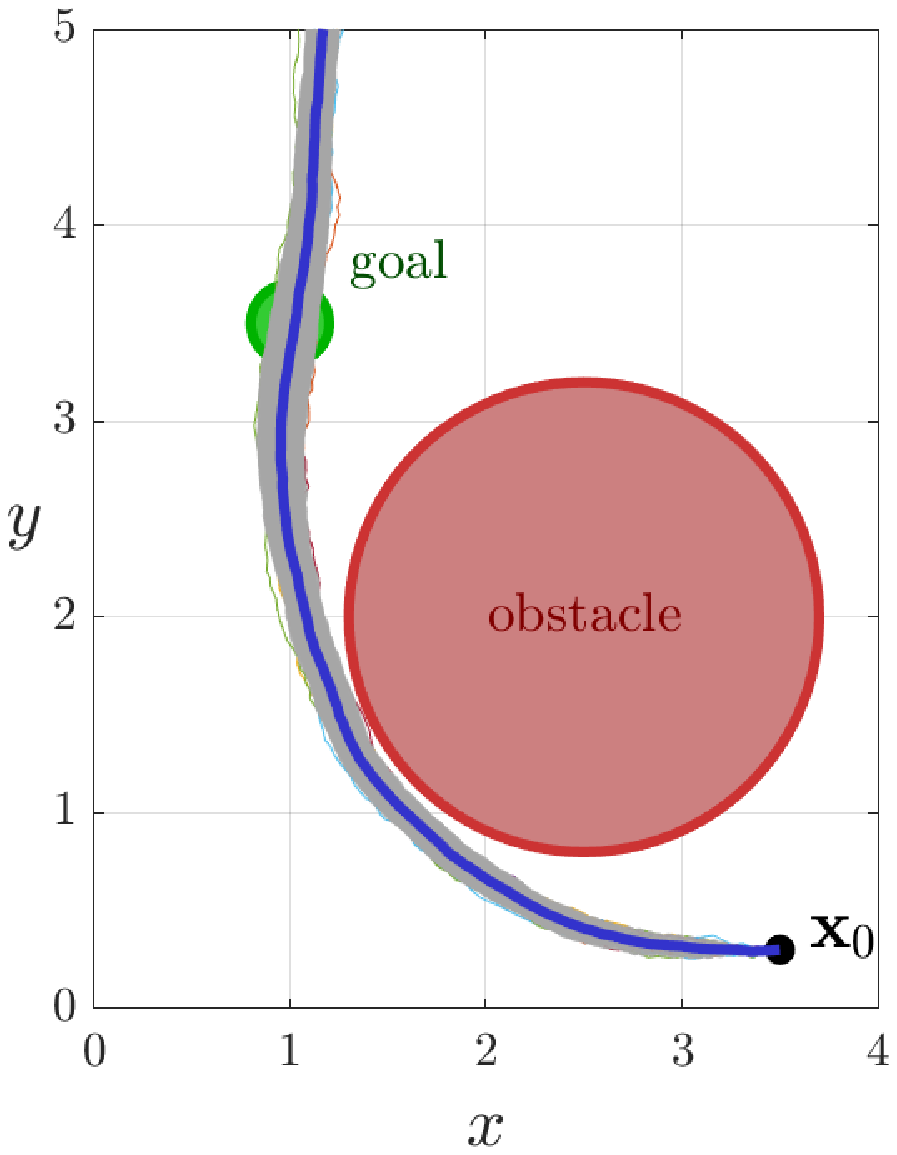}
			\caption{LIN: $C = 4.697 \pm 0.229$, \\ \phantom{(b) LIN::}  $\rho^{\varphi} = 0.104 \pm 0.070$}
		\end{subfigure}%
		\begin{subfigure}{0.5\linewidth}
			\centering
			\includegraphics[width=\linewidth,trim={2cm 0.2cm 2.4cm 0.05cm},clip]{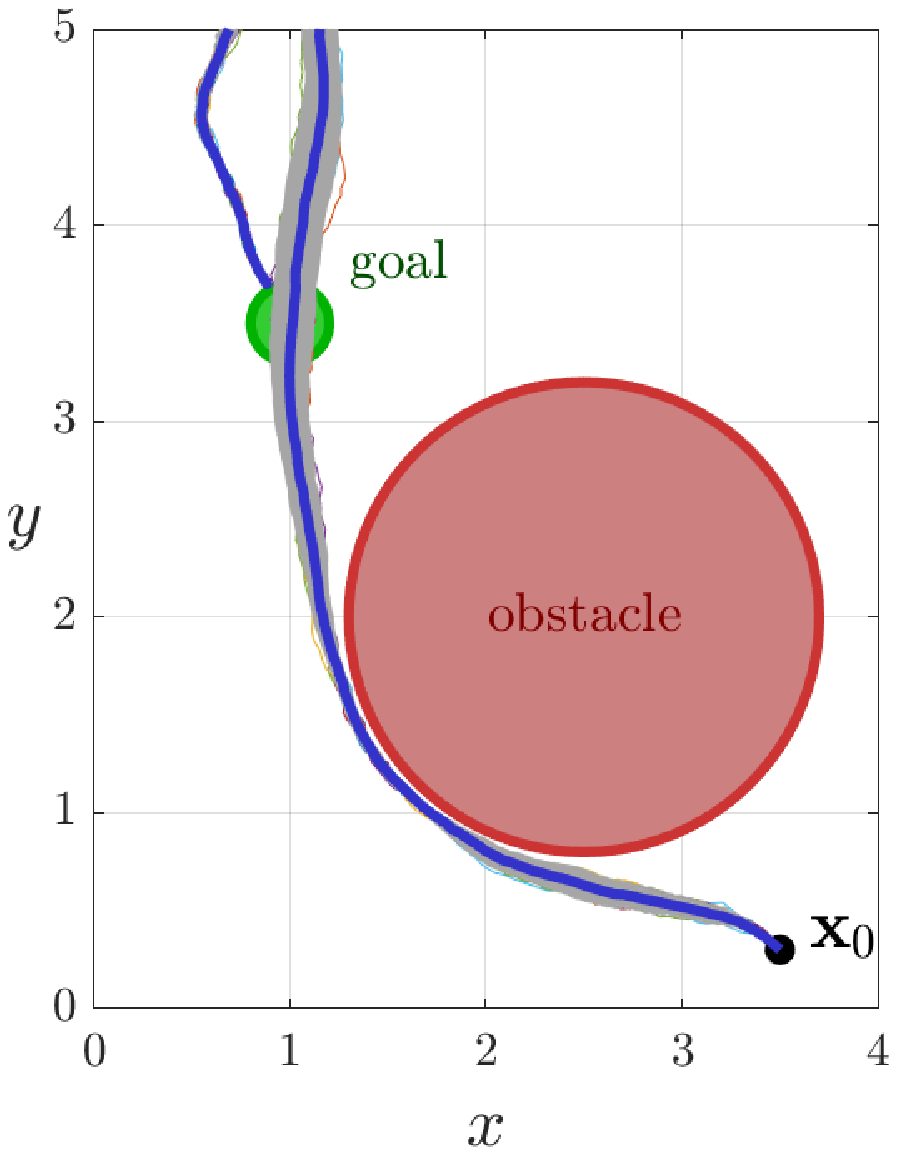}
			\caption{PPC: $C = 4.553 \pm 0.256$, \\ \phantom{(b) PPC::} $\rho^{\varphi} = 0.054 \pm 0.018$}
		\end{subfigure}
		\caption{Robustness of the \ac{PI2} algorithm variants with respect to system noise. The shaded gray area corresponds to 2 standard deviations of a Gaussian distribution fitted to 30 sample runs of the obtained controllers. With the PPC-based law, the optimal robustness is achieved with lower variance.}
		\label{fig:scenario1_noise}
	\end{minipage}
\end{figure}

Next, we include a disturbance $\w$ in the system dynamics with covariance $\SIG[w] = 0.2\I[2]$, 20\% of the input bound. The eliteness parameter $\epsilon$ was changed to 50\% and the penalty $\lambda$ was increased linearly to 50000 throughout $K=50$ iterations to achieve the desired minimal robustness measure even in the presence of noise. Furthermore, the Nesterov acceleration scheme was turned off for a better final result as it is known to amplify the effects of noise and thus hinder convergence in later iterations \cite{sutskever2013importance}. A sample result from both the \ac{PPC} and linear state feedback-based \ac{PI2} variants is depicted in Fig. \ref{fig:scenario1_noise}. The figure shows that the \ac{PPC} variant provides more robustness against noise due to the feedback nature of the base controller continuously correcting for its influence. The effect is most visible near the obstacle, and leads to a lower variation of the robustness measure, which in turn allows a more optimal solution in terms of the cost $C(\tau)$ to be found while aiming to keep $\rho^{\varphi} \ge \rho_{\min}$. Due to the prescribed funnel, the robot is also pushed back towards the goal region near the end of the simulated time frame.

%% file: results_complex.tex
\subsection{Complex task} \label{section:results_complex}

Consider a system of two ground vehicles and a drone described by (2D) single integrator dynamics and subject to the consensus protocol \cite{mesbahi2010graph} with additional free inputs:
\begin{equation}
\x<\dot>(t) = -0.1(\L \otimes \I[2])\x(t) + \u(t),
\end{equation}
which fits the system (\ref{eq:systemDynamics}) with the known input term $g(\x) = \I$ and unknown $f(\x) = -0.1(\L \otimes \I[2])$. The subscripts $\x[i]$ and $\u[i]$, $i = 1,2,3$, will refer to the location and inputs of the $i$-th robot. The input constraint is $\norm[2]{\u[i]} \le 1$ for each robot. The matrix $\L$ is the so-called Laplacian of the graph describing agent connections within the consensus protocol \cite{mesbahi2010graph}; assuming a complete graph it becomes 
\begin{equation}
\L = \bmat{2 & -1 & -1 \\ -1 & 2 & -1 \\ -1 & -1 & 2}. 
\end{equation}
The robots' initial locations are $\x[1,0] = [3.0\ 0.8]\tp$, $\x[2,0] = [2.0\ 0.8]\tp$, and $\x[3,0] = [1.2\ 0.7]\tp$.

The ground robots are tasked with reaching and staying within $r_g = 0.1$ meters of $\x[g1] = [2.0 \ 4.2]\tp$ and $\x[g2] = [3.0 \ 4.2]\tp$ within $7s$ while maintaining a mutual distance between $d_{12}^{\min} = 1 - \Delta d_{12}$ and $d_{12}^{\max} =1 + \Delta d_{12}$, $\Delta d_{12} = 0.1$. Furthermore, they must avoid a circular obstacle of radius 1m centered at $\x[o] = [2.5 \ 2.5]\tp$ by $r_o = 1.2$ during this maneuver (in order to leave space for, e.g., a carried object). The drone is tasked with reaching and staying within $r_a = 0.1$ meters from the middle of the two ground robots within 3 seconds. The goal is to satisfy this task with robustness $\rho_{\text{min}} = 0.02$ while minimizing the sum of each robot's extended energy, i.e., $C(\tau) = \sum_{i=1}^{3}\int_{0}^{T}\u[i]\tp \u[i]$.  The scenario is simulated for $T = 10s$ with resolution $\Delta t = 0.01s$.

\begin{figure}[hb]
	\centering
	\includegraphics[width=.662\linewidth,trim={0 0.7cm 0 0},clip]{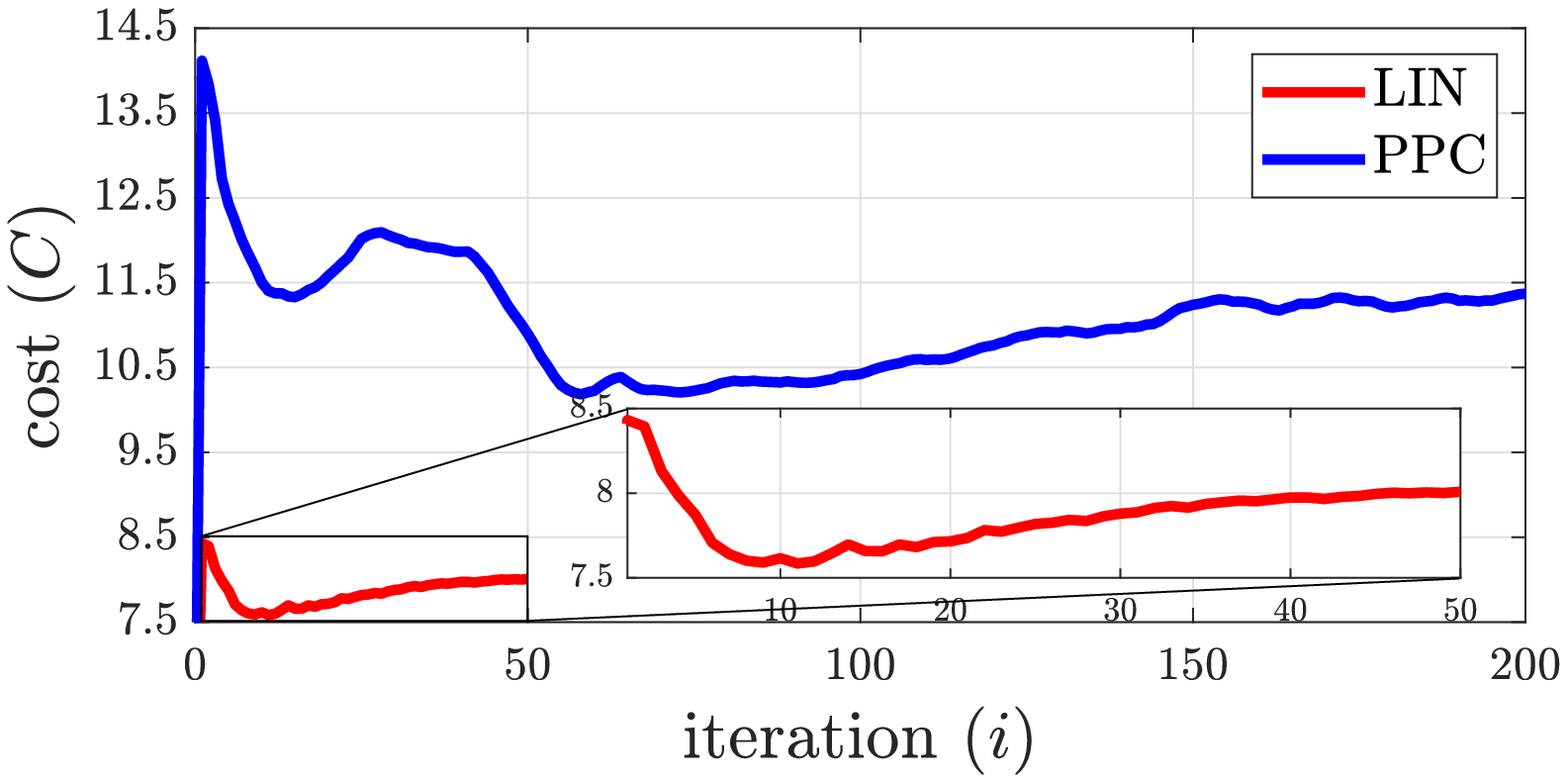}
	\includegraphics[width=.662\linewidth]{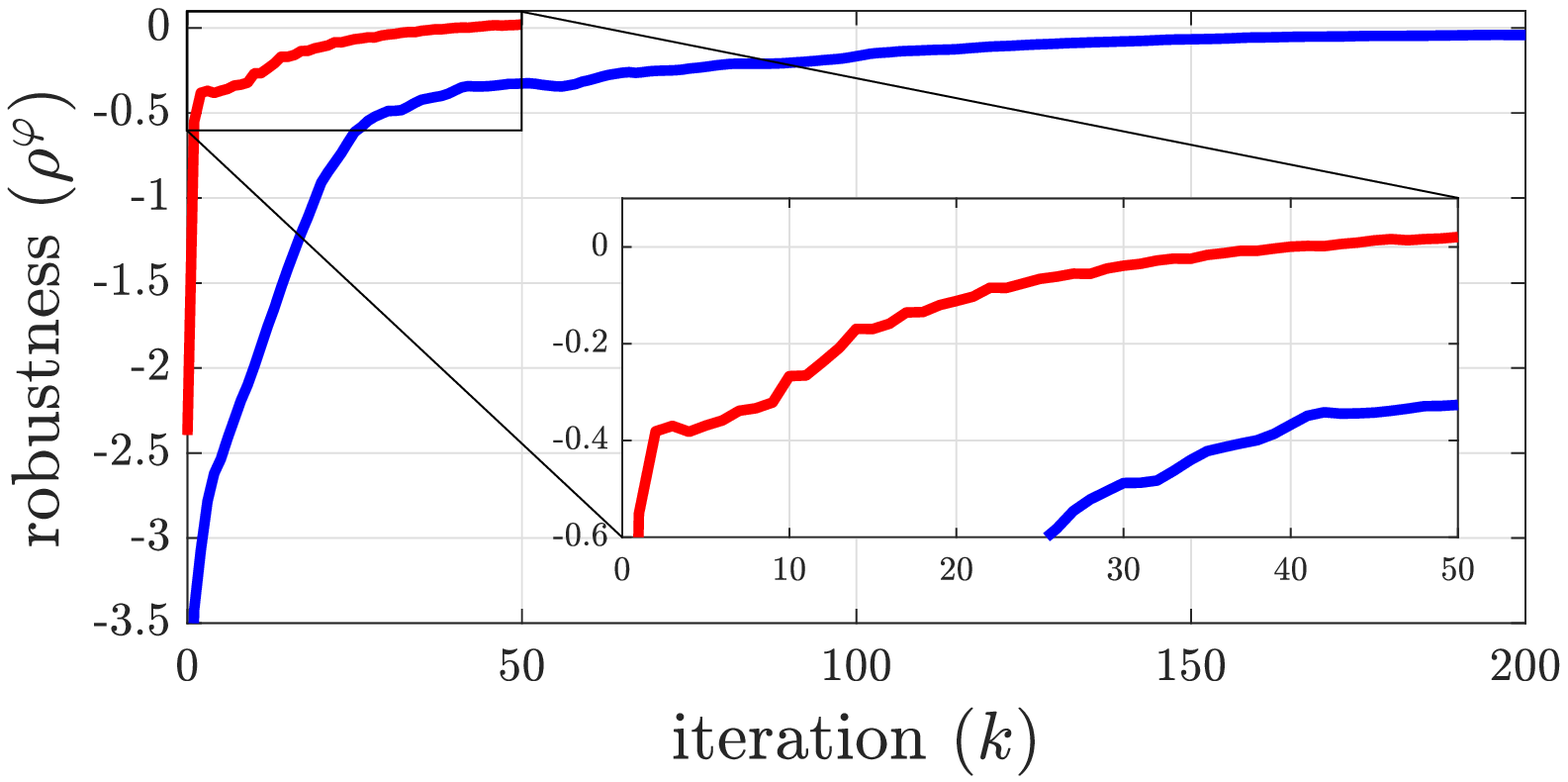}		
	\caption{Sample convergence of the cost $C(\tau)$ and robustness $\rho^{\varphi}$ during the solution for scenario 2 using Algorithm \ref{alg:pi2} in both its linear state feedback and \ac{PPC} guided forms.}
	\label{fig:scenario2_measures}
\end{figure}

\begin{figure*}[th]
	\centering
	\begin{subfigure}{0.33\textwidth}
		\centering
		\includegraphics[width=0.82\linewidth,trim={0.8cm 0.3cm 0.8cm 0.7cm},clip]{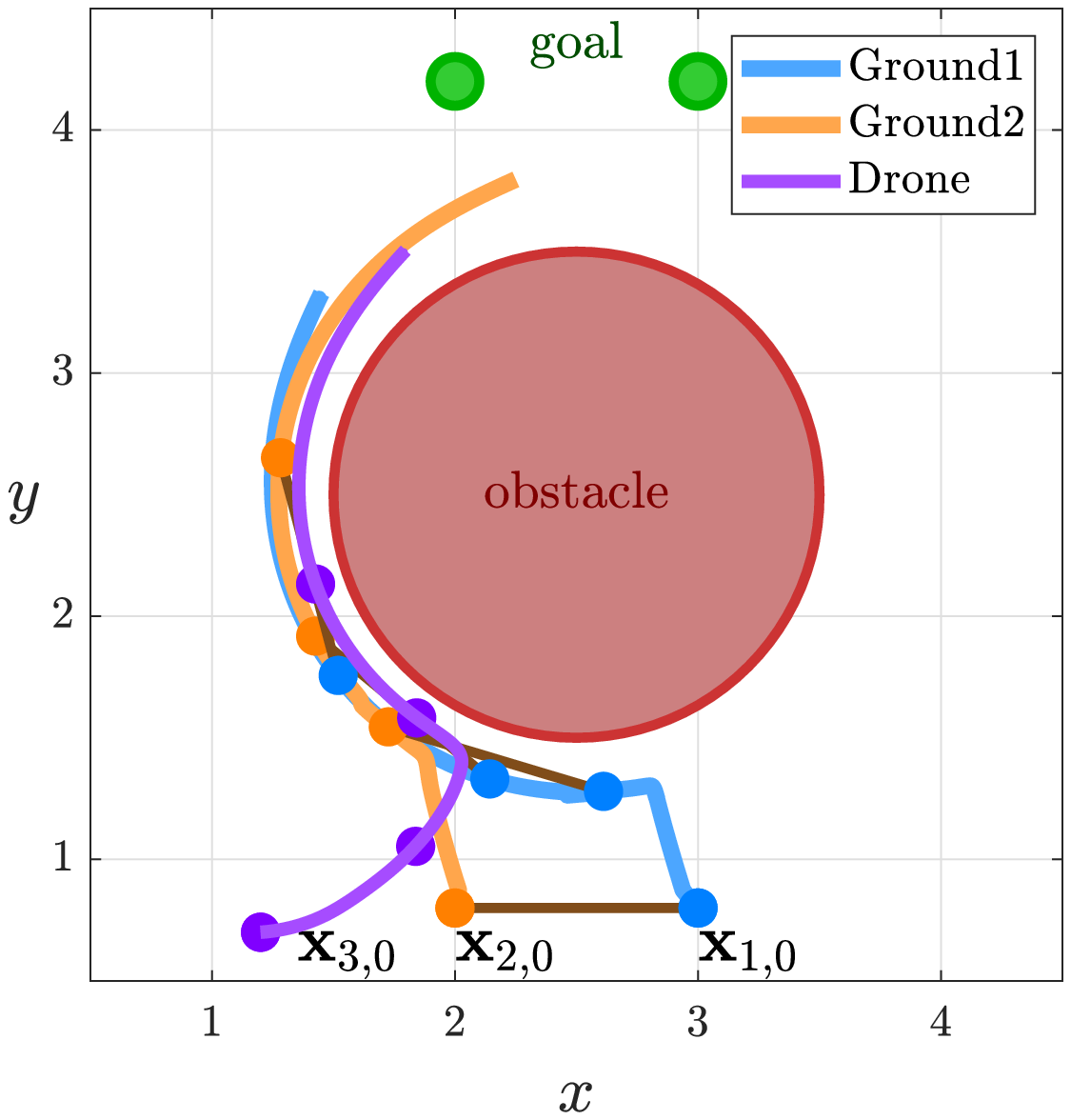}
		\caption{PPC(initial): $\rho^{\varphi} = -2.39$, $C = 4.87$}
	\end{subfigure}
	\begin{subfigure}{0.33\textwidth}
		\centering
		\includegraphics[width=0.82\linewidth,trim={0.8cm 0.3cm 0.8cm 0.7cm},clip]{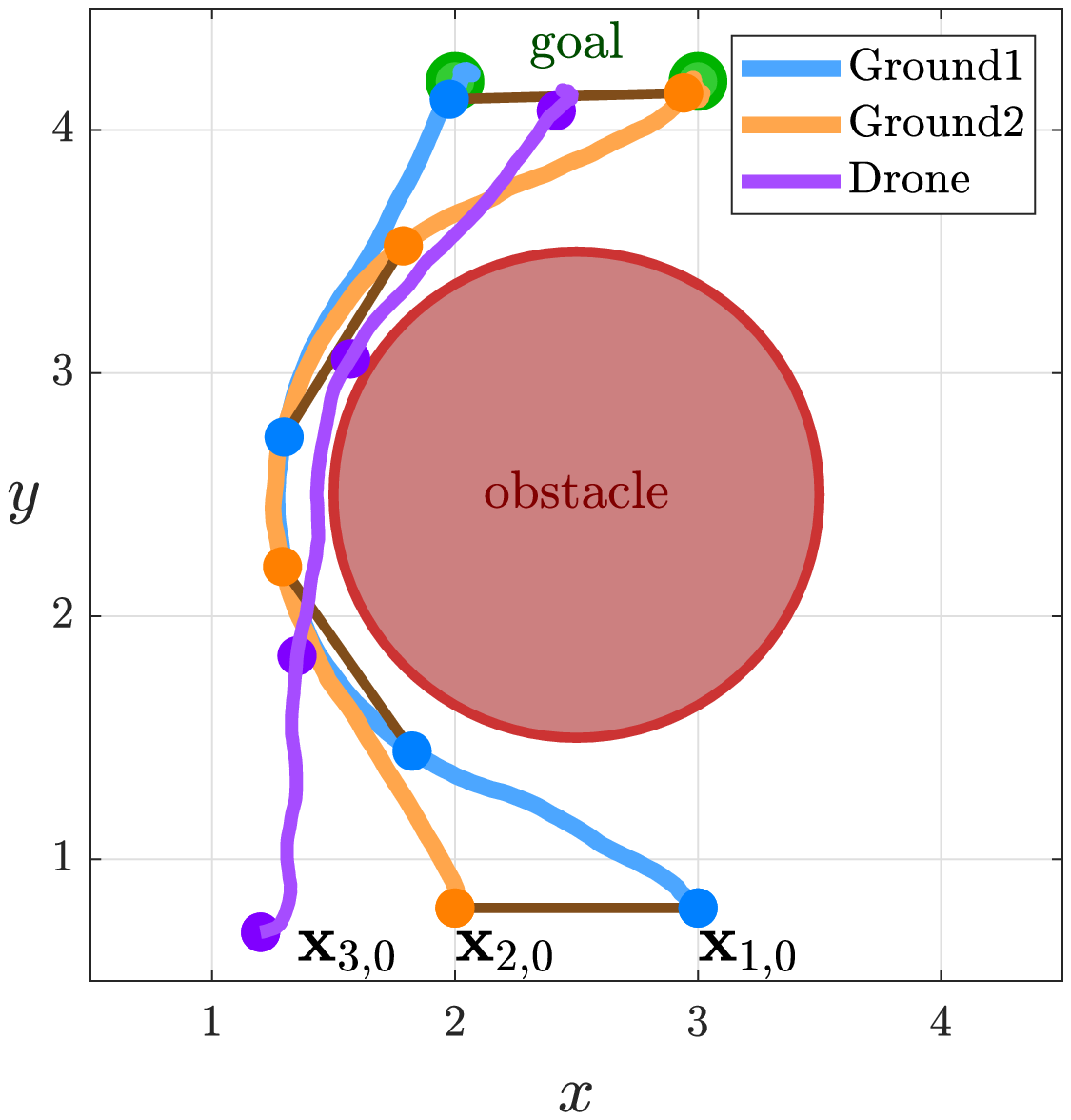}
		\caption{PPC(final): $\rho^{\varphi} = 0.020$, $C = 8.01$}
	\end{subfigure}%
	\begin{subfigure}{0.33\textwidth}
		\centering
		\includegraphics[width=0.82\linewidth,trim={0.8cm 0.3cm 0.8cm 0.7cm},clip]{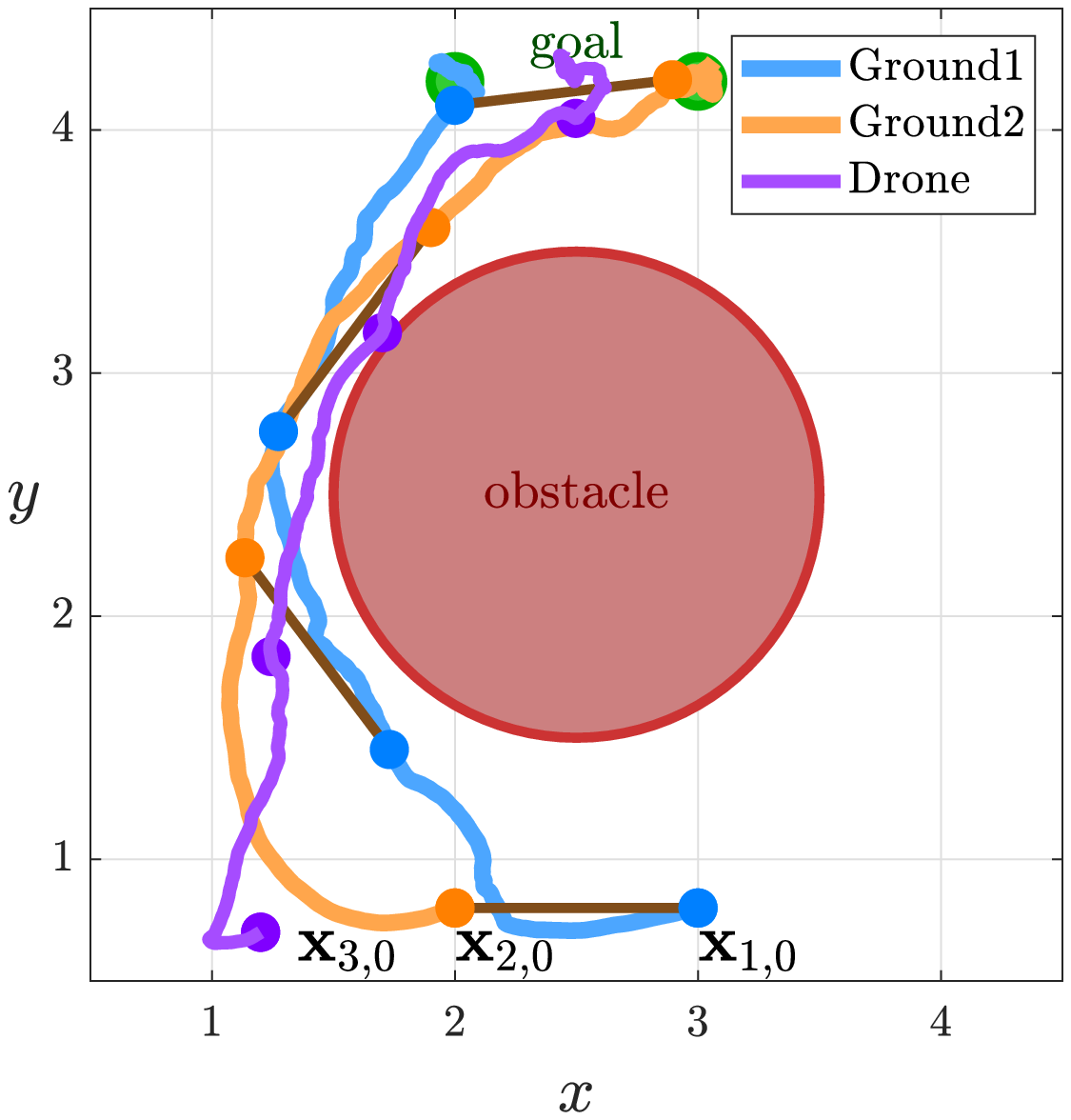}
		\caption{LIN: $\rho^{\varphi} = -0.041$, $C = 11.37$ }
		\label{fig:scenario2_sampleResultLIN}	
	\end{subfigure}%
	\caption{Sample solution to the complex scenario described in Section \ref{section:results_complex}. The location of the robots is shown at 4 evenly spaced points in time until the goal areas are reached around $t=7$s. Without a \ac{PPC} guide for task satisfaction, the algorithm has trouble effectively minimizing the cost $C(\tau)$ as seen from the raggedness of the resulting trajectories in (c).}
	\label{fig:scenario2_sampleResult}
\end{figure*}

A formal description of the task within the STL framework is given as follows. Define the non-temporal formulas $\psi_i = (\norm{\x[i] - \x[gi]} \le r_{g})$ for $i=1,2$, $\psi_3 = (\norm{\x[1] - \x[2]} \le d_{12}^{\max})$, $\psi_4 = (\norm{\x[1] - \x[2]} \ge d_{12}^{\min})$, $\psi_5 = (\norm{\x[1] - \x[o]} \ge r_{o})$, $\psi_6 = (\norm{\x[2] - \x[o]} \ge r_{o})$, and $\psi_7 = (\norm{(\x[1] + \x[2])/2 - \x[3]} \le r_{a})$. The corresponding temporal formulas are then $\phi_{i} = \eventually[0][7] \always[0][\infty] \psi_i$ for $i = 1,2$, $\phi_{i} = \always[0][\infty]\psi_i$ for $i = 3\dots 6$, and $\phi_{7} = \eventually[0][3]\always[0][\infty]\psi_7$. The full task specification is thus given as $\varphi = \bigwedge_{i=1}^{7} \phi_{i}$. The funnels aiming to enforce the subtasks are described as in the first scenario by the parameters $\rho_{\text{max}} = \{r_g, r_g, \Delta d_{12}, \Delta d_{12}, 1.0, 1.0, r_a\}$, $\gamma_0 = \{{-}4.0, {-}4.0, \rho_{\text{min}}, \rho_{\text{min}}, \rho_{\text{min}}, \rho_{\text{min}}, {-}2.0\}$, and finally $\gamma_{\infty} = \{\rho_{\text{min}}, \rho_{\text{min}}, \rho_{\text{min}}, \rho_{\text{min}}, \rho_{\text{min}}, \rho_{\text{min}}, \rho_{\text{min}}\}$, with the \mbox{$i$-th} element of each set corresponding to the values used for the $i$-th subtask. The transformation functions take the form of a linear-exponential function defined by the parameters $\beta = \{2, 2, 0.2, 0.2, 0.2, 0.2, 1\}$, $B = \{6, 6, 6, 6, 6, 6, 6\}$, and $\xi_c = \{0.5, 0.5, 0.8, 0.8, 0.8, 0.8, 0.8\}$. The hyperparameter values used in Algorithm \ref{alg:pi2} are $K = 50$, $N = 100$, $\epsilon = 80\%$, $\C[t,0] = 2\cdot10^{-4} \I$, and $\C[t, \text{min}] = 2\cdot10^{-7}\I$ for all time steps. The penalty term $\lambda$ was spaced logarithmically from $2$ to $10000$ throughout the $K$ iterations in order to enforce task satisfaction. The scenario is depicted in Fig. \ref{fig:scenario2_sampleResult} along with obtained sample results; the convergence behavior is shown on Fig. \ref{fig:scenario2_measures}. The LIN variant uses the feedback $\K[t] = \I[6]$.

Examining the resulting trajectories in detail, we can see that the distance traveled by the ground robots is minimized and their speed is such that the goals are reached at the latest possible $0.7$s in order to reduce the input efforts. The drone also maintains a more or less straight path near the middle of the two ground robots while lagging behind as much as possible to minimize its input effort. We note that without the \ac{PPC} base law as a guide, we were unable to tune the parameters for the LIN variant of \ac{PI2} to achieve task satisfaction with a remotely optimal cost. For the result shown in Fig. \ref{fig:scenario2_sampleResultLIN}, the initial exploration $\C[t,0]$ was increased tenfold and the algorithm ran for $K=200$ iterations.

%% file: conclusions.tex
\section{CONCLUSIONS} \label{section:conclusions}

In this work, we examined the possibility of using a \ac{PPC} base law to guide the \ac{PI2} reinforcement learning algorithm in order to solve optimal control problems involving \ac{STL} task specifications. The method offers multiple benefits, such as increased computational efficiency and robustness to noise and hyperparameters, as well as the ability to cope with more complicated task specifications. These advantages were illustrated in a simulation study of two sample scenarios. 

The results give incentive for developing \ac{STL} base laws that guarantee task satisfaction for a wider range of system dynamics and under increasingly complex task specifications. Further research possibilities include automating hyperparameter choices for the proposed algorithm, as well as extending the method to the multi-agent domain by decentralizing the base control and learning aspects. We also note that policy improvement constitutes an intermediate step within policy search \cite{chebotar2017path}, which finds general policies for arbitrary initial conditions of the system. In order to get the computational benefits of offline training from \acl{RL}, extending the presented results to such a policy search framework is also of considerable interest.